\documentclass[]{emulateapj}
\usepackage{epsfig, natbib, graphicx, color}
\usepackage{apjfonts}

\input epsf

\newcommand{\avg}[1]{\left\langle #1 \right\rangle}

\newcommand{\kpc}{\mbox{kpc}}

\newcommand{\msun}{M_\odot}
\newcommand{\bm}[1]{\mathbf{#1}}

\newcommand{\LCDM}{\Lambda\mbox{CDM}}

\newcommand{\los}{\bm{\hat n}}
\newcommand{\squad}{\sigma^{(4)}}
\newcommand{\sdouble}{\sigma^{(2)}}

\newcommand{\pquad}{p^{(4)}(\mu)}

\shortauthors{ROZO ET AL}
\shorttitle{Lensing Biasing of Triaxial Halos}

\begin{document}
\title{Biases in the Gravitational Lens Population Induced by Halo and Galaxy Triaxiality}
\author{Eduardo Rozo\altaffilmark{1},
Jacqueline Chen\altaffilmark{2},
Andrew R. Zentner\altaffilmark{3,4,5}
}

\altaffiltext{1}{Center for Cosmology and Astro-Particle Physics (CCAPP), The Ohio State University, Columbus, OH, USA}
\altaffiltext{2}{Argelander-Institut f\"{u}r Astronomie, University of Bonn, Auf dem H\"{u}gel 71, 53121 Bonn, Germany}
\altaffiltext{3}{Department of Physics and Astronomy, University of Pittsburgh, Pittsburgh, PA 15260, USA}
\altaffiltext{4}{Kavli Institude for Cosmological Physics and Department of Astronomy, Chicago, IL 60637, USA}
\altaffiltext{5}{The Enrico Fermi Institute, The University of Chicago, Chicago, IL 60637, USA}

\begin{abstract}
The lensing cross section of triaxial halos depends on the relative orientation between a halo's 
principal axes
and its line of sight.  Consequently, a lensing subsample of randomly oriented halos is not, in general,
randomly oriented.  
Using an isothermal mass model for the lensing galaxies and their host halos, we show 
that the lensing subsample of halos that produces doubles is preferentially aligned along the
lines of sight, whereas halos that produce quads tend to be projected along their middle axes.
These preferred orientations result in different projected ellipticity
distributions for quad, doubles, and random galaxies.   We show that $\approx 300$ lens systems 
must be
discovered to detect this effect at the $95\%$ confidence level.  We also investigate the importance
of halo shape for predicting the quad-to-double ratio and find that the latter depends quite sensitively
on the distribution of the short-to-long axis ratio, but is otherwise nearly independent of halo shape.
Finally, we estimate the impact of the preferred orientation 
of lensing
galaxies on their projected substructure mass fraction, and find that the observed alignment
between the substructure distribution and the mass distribution of halos result in a negligible bias.
\end{abstract}
 \keywords{
galaxies, halos, lensing
}

\section{Introduction}

Statistics of lensing galaxies have been used as cosmological and galaxy formation probes since early
in the modern history of gravitational lensing \citep[][]{turneretal84}.  Lensing rates can
be used to constrain dark energy \citep[][]{fukugitaetal92, chae03,mitchelletal05,
chae07,ogurietal07}, to probe the structure of lensing galaxies \citep[][]{keeton01d,kochanekwhite01,
chae05},
and to probe galaxy evolution \citep[][]{chaemao03, ofeketal03,rusinkochanek05}.   While the
use of lensing statistics as a cosmological probe has had mixed success, particularly early on,
it remains a unique probe with entirely different systematics from more traditional approaches.
Consequently, lensing statistics are likely to remain a fundamental cross-check
of our understanding of cosmology and galaxy evolution.

One of the difficulties that confronts the study of lensing statistics is that, in general,
the halo population that produces gravitational lenses can in fact be a highly biased
subsample of the general halo population.  For instance, it has long been known
that while early type galaxies compose only $\approx 30\%$ of all luminous galaxies,
the majority of lensing galaxies are in fact early type since these tend to be more massive
and reside in more massive halos than their late counterparts.  By the same token,
lensing early type galaxies tend to have higher luminosity and velocity dispersions
than non-lensing early type galaxies \citep[][]{moelleretal06, boltonetal06}.  Overall, then, when interpreting
lensing statistics, one ought to always remember that by selecting lensing galaxies one is 
automatically introducing an important selection effect that can significantly bias the 
distribution of any galaxy observable that has an impact on the lensing probabilities.  
Here, we consider one such source of bias, the triaxiality of galaxy halos.\footnote{Throughout
this work, we will be using the term galaxy and halo more or less interchangeably.  The reason
for this is that we are primarily focused on the impact of halo triaxiality on the lensing
cross section, and the latter depends only on the {\it total} matter density.  Consequently, differentiating
between halo and galaxy would only obfuscate presentation and introduce unnecessary
difficulties.  For instance, while modeling the total matter distribution as
isothermal is a reasonable approximation, neither the baryons nor the dark matter by itself 
is isothermally distributed.  Thus,
it is much simpler to adopt an isothermal model, and refer to the baryons plus dark
matter as a single entity, than to try to differentiate between the two.  Likewise, when discussing
triaxiality, what is important in this work is the triaxiality of the total matter distribution.}

That halo triaxiality can have important consequences for
lensing statistics has been known for several years.  
For instance, \citet[][]{ogurikeeton04} have shown that triaxiality can significantly
enhance the optical depth of large image separation lenses.  Similar conclusions have
been reached concerning the formation of giant arcs by lensing clusters 
\citep[see e.g.][and references therein]{ogurietal03, rozoetal06c, hennawietal07}. 
Curiously, however, little 
effort has gone into investigating
how observational properties of lensing galaxies can be different from those 
of the galaxy population as a whole due to the triaxial structure of galactic halos.
This work addresses this omission.

The first observable we consider is the projected axis ratio of lensing galaxies.  
Roughly speaking, given that non-zero ellipticities are needed in order to produce
quad systems, one would generically expect lenses that lead to this image configuration
to be more elliptical than the overall galaxy population.
Likewise, lensing galaxies that produce
doubles should, on average, be slightly more circular than a random galaxy.   
There can, however, be complications for these simple predictions due to
halo triaxiality.   For instance, given a prolate
halo, projections along the long axis of the lens will result in highly concentrated,
very circular profiles.  Will the increase in Einstein radius of such projections 
compensate for the
lower ellipticity of the system, implying most quads will be projected along their
long axis, or will it be the other way around?  
Clearly, the relation between ellipticity and lensing cross sections is
not straightforward once triaxiality of the lensing galaxies is taken into 
account, but it seems clear that there should be some observable difference
between the ellipticity distribution of lensing galaxies and that of all early types.
Interestingly, no such difference has been observed \citep[][]{keetonetal97,rusintegmark01},
which seems to fly in the face of our expectations \citep[though see also the discussion in][]{keetonetal98}.  
Is this actually a problem, or
will a quantitative analysis show that the consistency of the two distributions
is to be expected?  Here, we explicitly resolve this question, and demonstrate that
current lens samples are much too small to detect the expected differences.

Having considered the ellipticity distribution of random and lensing galaxies, it is then a natural
step to investigate the impact of halo triaxiality on predictions of the quad-to-double
ratio.  Specifically, it is well known that the quad-to-double ratio is sensitive to the ellipticity
distribution of lensing galaxies \citep[][]{keetonetal97}, so if lensing
can bias the distribution of ellipticities in lensing galaxies, then it should also affect the
predicted quad-to-double ratios.  This is an important point because it has been argued that
current predictions for the quad-to-double ratio are at odds with observations.
More specifically, the predicted 
quad-to-double ratio for the CLASS \citep[Cosmic Lens All-Sky Survey,][]{myersetal03,browneetal03} 
sample of gravitational lenses is too low relative to
observations \citep[][]{rusintegmark01,hutereretal05}.  
Curiously, however, recent work on the quad-to-double ratio observed in the
SQLS \citep[Sloan Digital Sky Survey Quasar Lens Search,][]{ogurietal06,inadaetal07}.
suggests that the exact opposite is true for the latter sample, namely, theoretical expectations
are too high relative to observations \citep[][]{oguri07}.   In either case, it is of interest
to determine how exactly does triaxiality affects theoretical predictions, especially since
the aforementioned difficulties with the CLASS sample has led various
authors to offer possibilities as to how one might boost the expected
quad-to-double ratios.  Specifically, one can boost the quad-to-double ration in the 
class sample either from the effect of 
massive satellite galaxies near the lensing galaxies \citep[][]{cohnkochanek04}, or through the 
large-scale environment of the lensing galaxy \citep[][]{keetonzabludoff04}.  Clearly,
we should determine whether halo triaxiality can be added to this list.

This brings us then to the final problem we consider here, namely whether the substructure population
of lensing galaxies is different from that of non-lensing galaxies. Specifically,
we have argued that lensing galaxies will not be isotropically distributed
in space.  Since the substructure distribution of a dark matter halo is typically aligned with its parent
halo's long axis \citep[][]{zentneretal05,libeskindetal05,agustssonbrainerd06,azzaroetal06}, 
it follows that the projected distribution of substructures for lensing galaxies may in fact
be different for lensing halos than for non-lensing halos.  Such an effect could be quite important given
the claimed tension between the Cold Dark Matter (CDM) predictions for the substructure mass
fraction of halos \citep[see][]{maoetal04} and their observed values \citep[][]{dalalkochanek02a,kochanekdalal04}.
Likewise, such a bias would impact the predictions for the level of astrometric and flux perturbations produced
by dark matter substructures in gravitational lenses \citep[][]{rozoetal06,chenetal07}.  Here, we wish to
estimate the level at which the projected substructure mass fraction of lensing halos could be affected
due to lensing biasing.

The paper is organized as follows: in section \ref{sec:biases} we derive the basic equations needed
to compute how observable quantities will be biased in lensing galaxy samples due to halo
triaxiality.  Section \ref{sec:model} presents the model used in this work to quantitatively estimate
the level of these biases, and discusses how lensing halos are oriented relative to the line
of sight as a function of the halos' axes ratios.  Section \ref{sec:axis} investigates the projected
axis ratio distributions of lensing versus non-lensing galaxies, and demonstrates that 
present day lensing samples are too small to detect the triaxiality induced biases we have 
predicted.  Section \ref{sec:ratio} discusses the problem of the quad to double ratio, and
section \ref{sec:subs} demonstrates that halo triaxiality biases the projected substructure
mass fraction in lensing halos by a negligible amount.  Section \ref{sec:caveats} discusses
a few of the effects we have ignored in our work and how these may alter our results,
and finally section \ref{sec:summary} summarizes our work and presents our conclusions.


\section{Lens Biases Induced by Triaxiality}
\label{sec:biases}

We begin by deriving the basic expressions on which we rely to estimate
the effects of halo triaxiality on the observed properties of lensing galaxies.
In particular, we show that since the lensing cross section for triaxial lenses
is in general not spherically symmetric, this implies that a population of randomly
oriented halos produces a non-random lens population.  Finally, we show that
the induced non-randomness of the lensing halo population can alter the mean
observational properties of these halos relative to the general halo population.

\subsection{The Lensing Cross Section}
\label{sec:cs}

Let $\bm{p}$ be a set of parameters that characterizes the projected 
gravitational potential of a halo.  For instance, $\bm{p}$ can be the Einstein radius of the
lens, its ellipticity, and so on.  Given a background source density $n_s(z_s)$ and a halo 
density $n_h(\bm{p},z_h)$, and in 
the absence of a flux limit, the mean number of lensing events per unit redshift per area
is given by
\begin{equation}
\frac{dN_{lenses}}{dz_sdz_hd\Omega} = n_s(z_s) n_h(\bm{p},z_h)
\frac{d\chi}{dz_s}\frac{d\chi}{dz_h}\sigma(\bm{p},z_h,z_s)
\end{equation}
where $\chi$ is the comoving distance to the appropriate halo or source redshift, and
\begin{equation}
\sigma(\bm{p},z_h,z_s) = \int_{lensing} d^2\bm{y}.Ä
\label{eq:cs}
\end{equation}
The integral is over all regions of the source plane that produce lensed images
of interest.  For instance, if one were interested in quadruply imaged sources,
the integral would be over all source positions that result in four image lenses.
The quantity $\sigma$ is called the \it lensing cross section, \rm  and of
particular interest to us will be the cross sections $\sigma^{(N)}$ for
producing $N$-image systems.

In reality, one always has some flux limit $F_{min}$ which corresponds
to a minimum source luminosity $L_{min}$.   Fortunately, the above argument
is easily generalized: let $dn_s(L,z_s)/dL$ be the number density of background 
sources with luminosity $L$.  Then, the mean number of lensing events becomes
\begin{equation}
\frac{dN_{lenses}}{dz_sdz_hd\Omega} = n_h\frac{d\chi}{dz_s}\frac{d\chi}{dz_h}
	\int d^2\bm{y} \int_{L_{min}/\mu(\bm{y})}^\infty dL\ \frac{dn_s(L,z_s)}{dL}.
\end{equation}
If the source luminosity function can be approximated by a power law 
$dn_s(L,z_s)/dL = AL^{-\alpha}$ (note both $A$ and $\alpha$ can depend on $z_s$),
the above expression reduces to 
\begin{equation}
\frac{dN_{lenses}}{dz_sdz_hd\Omega} = n_b(>L_{min}) n_h 
	\frac{d\chi}{dz_s} \frac{d\chi}{dz_h} \sigma_B(\bm{p},\alpha,z_h,z_s)
\label{eq:bcs}
\end{equation}
 where $n_b(>L_{min},z_s)$ is the number density of sources above the flux limit
 \it in the absence of lensing, \rm  and $\sigma_B$ is given by
 \begin{equation}
 \sigma_B (\bm{p},z_h,z_s,\alpha)= \int d^2\bm{y}\ \mu(\bm{y})^{\alpha-1}
 \end{equation}
where $\mu(\bm{y})$ is the total magnification of a source at position $\bm{y}$.
Following \citet[][]{hutereretal05}, we call $\sigma_B$ the \it biased cross
section. \rm  Indeed, since the distribution of magnifications $p(\mu)$
among all lensing events is given by
\begin{equation}
p(\mu) = \frac{1}{\sigma}\int d^2\bm{y}\ \delta(\mu(\bm{y})-\mu)
\end{equation}
where $\sigma$ is the (unbiased) lensing cross section defined in Eq.
\ref{eq:cs}, then we can rewrite Eq. \ref{eq:bcs}   as
\begin{equation}
\sigma_B = \avg{\mu^{\alpha-1}}\sigma,
\end{equation}
where
\begin{equation}
\avg{\mu^{\alpha-1}} = \int d\mu\ p(\mu)\mu^{\alpha-1}.
\end{equation}
Thus, the net effect of gravitational magnification on the frequency of lensing events 
can be summarized as a biasing factor $\avg{\mu^{\alpha-1}}$ that multiplies the 
unbiased lensing cross section $\sigma$.

\subsection{Triaxiality and Lensing Biasing}

Let $\bm{P}$ characterize the mass distribution of a triaxial halo, and let $\bm{\hat n}$
be the orientation of the halo's long axis relative to the line of sight.  The halo's two dimensional
potential is then characterized by a new set of parameters $\bm{p}(\bm{P},\bm{\hat n})$
which depend on the halo properties $\bm{P}$ and the particular line of sight $\los$ along
which the halo is being viewed.  For instance, the vector $\bm{P}$ can include such halo properties
as halo mass and axis ratios, whereas $\bm{p}$ could include parameters such as the Einstein
radius of the projected mass distribution as well as the projected axis ratio.

As discussed above, the mean number of lensing events per unit redshift by a
halo along a given line of sight is given by Eq. \ref{eq:bcs}.  For convenience, we
define the halo and source surface densities
$d\Sigma_h/d\bm{P}$ and $d\Sigma_s/dz_s$ via
\begin{eqnarray}
\frac{d\Sigma_h}{d\bm{P}dz_h}\ & =\ & \frac{dn_h}{d\bm{P}}\frac{d\chi}{dz_h} \\
\frac{d\Sigma_s}{dz_s} \ & =\ & n_s(>L_{min})\frac{d\chi}{dz_s}.
\end{eqnarray}
In terms of these surface densities, and assuming a randomly-oriented distribution of
halos, the mean number of lenses per unit area as a function of their orientation 
$\los$ is given by
\begin{equation}
\frac{dN_{lenses}}{d\bm{P}d\los dz_sdz_hd\Omega} = \frac{1}{2\pi}\frac{d\Sigma_s}{dz_s}\frac{d\Sigma_h}{d\bm{P}dz_h}
	\sigma_B(\bm{p}(\bm{P},\los),z_h,z_s,\alpha).
\label{eq:numlens}
\end{equation}
The prefactor of $1/(2\pi)$ arises from the fact that 
$dn_h/d\bm{P}d\los = (dn_h/d\bm{P})/(2\pi)$ due to our assumption of randomly
oriented halos.\footnote{If $\los$ denotes the angle between the line of sight and a
specified halo axis, and given that $\los$ and $-\los$ correspond to the same line of
sight, then it is evident that the space of all lines of sight is simply $S^2/Z_2$ - a sphere
with its diametrically opposed points identified.  The volume of such a space with the usual
metric is thus simply $2\pi$.}  We emphasize that Eq. \ref{eq:numlens} characterizes the number of
lenses \it as a function of the relative orientation $\los$ between the halo's major
axis and the line of sight. \rm  Thus, to compute the total number of lenses irrespective
of halo orientation, we would simply integrate the above expression over all lines of 
sight $\los$.  

There is an absolutely key point to be made concerning Eq. \ref{eq:numlens},
which provides the motivation behind this work.  Specifically, we note that the
number of lenses is proportional to $\sigma_B(\bm{p}(\bm{P},\los))$.  This implies
that even though the overall halo population does not have a preferred orientation
in space, {\it the lens population is not randomly oriented}, a fact which can
have observable consequences.    
In particular, given an observable halo property $f(\bm{P},\los)$ that depends on the
line of sight projection (e.g. the projected
axis ratio or projected substructure mass fraction), the mean value of $f$ over all 
$\bm{P}$ halos is simply
\begin{equation}
\avg{f|\bm{P}}_{halos} = \int \frac{d^2\los}{2\pi} f(\bm{P},\los),
\end{equation}
whereas the mean value of $f$ over all lenses is given by
\begin{equation}
\avg{f|\bm{P}}_{lenses} = \frac{1}{\avg{\sigma_B|\bm{P}}}
	\int \frac{d^2\los}{2\pi} \sigma_B(\bm{p}(\bm{P},\los)) f(\bm{P},\los)
\label{eq:losdist}
\end{equation}
where $\avg{\sigma_B}$ is the average value of $\sigma_B$ over all lines of sight,
\begin{equation}
\avg{\sigma_B|\bm{P}} = \int \frac{d^2\los}{2\pi} \sigma_B(\bm{p}(\bm{P},\los)).
\end{equation}
Thus, in general, one expects that the mean value of $f$ over all lenses and over all halos
will be different.  In the next few sections, we identify a few halo properties that depend on line
of sight projection, and determine whether lensing biases induced by triaxiality are likely
to be significant.


\section{The Model}
\label{sec:model}

We estimate the impact of halo triaxiality on the properties of lenses by considering a 
triaxial isothermal profile.  The merit of this approach is its simplicity: because of the simple
form of the matter distribution in this model, we can compute all of the relevant quantities
in a semi-analytic fashion, and the main
features of the model can be easily understood, thereby providing an important reference point 
for investigating more elaborate models.   Moreover, by working out in detail a simple analytic
model, our results provide an ideal test bed for more involved numerical codes, which would then
allow us to investigate how our conclusions are changed as more complicated models are allowed
(Chen et al. 2007, in preparation).

\subsection{Semi-Analytical Modeling}

Our analytical halo model is that of a simple triaxial isothermal profile of the form
\begin{equation}
\rho(\bar\bm{x}) = N(q_1,q_2)\frac{\sigma_v^2}{2\pi G}
	\frac{1}{x^2/q_1^2+y^2+z^2/q_2^2}
\label{eq:3dsie}
\end{equation}
where $q_1$ and $q_2$ are the axis ratios of the profile and we have chosen a coordinate
system that is aligned with the halo's principal axes, and such that 
$1\geq q_1 \geq q_2$.\footnote{i.e. $q_1$ is the ratio of medium to long
axis of the halo, whereas $q_2$ is the ratio of the short to long axis.  The motivation behind our
particular choice of axis labeling will
be made clear momentarily.}   The normalization 
constant $N(q_1,q_2)$ is chosen to ensure that the mass contained within a sphere of
radius $r$ be independent of the axis ratios for fixed velocity dispersion $\sigma_v^2$, the latter
being the velocity dispersion of the Singular Isothermal Sphere (SIS) obtained when $q_1=q_2=1$.

Let then $(\theta,\phi)$ denote a line of sight.  In appendix \ref{app:proj}, we show that the
corresponding projected surface mass density profile is that of a Singular Isothermal 
Ellipsoid (SIE) which, following \citet[][]{kormannetal94}, we write as
\begin{equation}
\Sigma(\tilde x, \tilde y) = \frac{\sqrt{q}\tilde \sigma_v^2}{2G}\frac{1}{\tilde x^2+q^2\tilde y^2}
\end{equation}
where both $q$ and $\tilde \sigma_v$ are known functions of $q_1,\ q_2$ and, in the case
of $\tilde \sigma_v$, of $N(q_1,q_2)\sigma_v^2$ (see Appendix
\ref{app:proj} for details).  In the above expression, $\tilde \sigma$ and $q$ are the effective velocity
dispersion and axis ratio respectively of the projected SIE profile.  As shown by \citet[][]{kormannetal94},
the lensing cross section for an SIE scales trivially with the Einstein radius $b$\footnote{By trivially,
we mean $\sigma\propto b^2$.}
\begin{equation}
b= 4\pi\frac{\tilde\sigma_v^2}{c^2}\frac{D_lD_{ls}}{D_s}
\label{eq:erad}
\end{equation}
of the profile.  Consequently, the distribution of halo orientations for a lens sample, 
$\rho(\los)=\sigma_B(\los)/\avg{\sigma_B}$,
is independent of the velocity dispersion $\sigma_v$ of the halo. 

There is one last important element of the model that needs to be specified, namely the luminosity
function of the sources being lensed.   Here, we take the luminosity function to be a power law with
slope of $-2$, which, while not exactly correct, is reasonably close to the slope of the luminosity
function of CLASS lenses \citep[][]{chae03,mckeanetal07}.    Moreover, this choice is ideally suited for numerical
work since in such a case the biased cross section is simply $\sigma_B=\avg{\mu}\sigma$,
implying that the biased cross section can be easily computed through uniform Monte Carlo sampling
of the image plane.
Since one of our goals in
this work is to provide a test case for more complicated numerical algorithms, we choose $\alpha=-2$.

Having fully specified our model, we can now easily compute the biased lensing cross section
for halos of any shape as a function of line of sight.  
Briefly, we proceed as follows.  First, 
we compute the biased lensing cross section for SIE profiles as a function of
the projected axis ratio $q$ for a grid of $q$ values.  These data points are then fit
using a third order polynomial fit, which we find is accurate to $\lesssim 1\%$.   Using
this simple fit for $\sigma_b(q)$, and the fact that we can analytically compute the 
Einstein radius and projected axis ratio for a triaxial halo along any line of sight,
we can readily compute the mean lensing cross section of a halo averaged over all lines
of sight.  For a detailed description of our calculations, we refer
the reader to the Appendices.

Before we end, however, 
it is important to remark here that, despite its simplicity, we expect our model is more than adequate
to investigate the qualitative trends that we would expect to observe in the data, and for providing
order of magnitude estimates of the impact of triaxiality.  Specifically, elliptical isothermal profiles appear 
to be excellent approximations
to the true matter distribution in real lens systems \citep[see e.g.][]{gerhardetal01, rusinma01, rusinetal03, 
rusinkochanek05, treuetal06, koopmansetal06, gavazzietal07}, so the triaxial 
isothermal mass distribution
considered here should provide a reasonably realistic model for order of magnitude estimates.   While
more sophisticated models are certainly possible \citep[see e.g.][]{jiangkochanek07}, it 
is our view that the simplicity of the isothermal model 
more than justifies our choice of profile for a first pass at the problem.


\subsection{The Distribution of Halo Orientations for Triaxial Isothermal Profiles}

Before we look at the distribution of halo orientations,
it is worth taking a minute to orient ourselves in the coordinate system we have chosen.  Consider first Eq. \ref{eq:3dsie}.
The distance from the center of the halo to the intercept of a constant density contour is
maximized for the $y$ axis, and minimized for the $z$ axis, while the $x$ axis is intermediate between the two.  If we 
then parameterize the line of sight using the circular 
coordinates $\theta$ and $\phi$ where $\theta$ is the angle with the $z$ axis and $\phi$ is the projected angle with 
the $x$ axis, then our coordinate system is such that it has the following properties.
\begin{itemize}
\item The $x, y,$ and $z$ axis of our coordinate system correspond to the middle, long, and short axis of the halo 
respectively.
\item Projections along $\cos(\theta)=1$ are along the short axis of the halo. 
\item Projections along $\cos(\theta)=0,\ \phi=0$ are along the middle axis of the halo.
\item Projections along $\cos(\theta)=0,\ \phi=\pi/2$ are along the long axis of the halo.
\end{itemize}
The nice thing about this particular choice of coordinates is that in the $\cos(\theta)-\phi$ plane, both the long
and the middle axis are represented by a single point, whereas the short axis is represented by an entire line.
As we shall see, projections along the middle and long axis maximize the lensing cross section of a halo 
for quad and double lenses respectively, so having that maximum be a single point in the space of lines of 
sight is a desirable quality of our chosen coordinate system.


\begin{figure}[t]
\epsscale{1.2}
\plotone{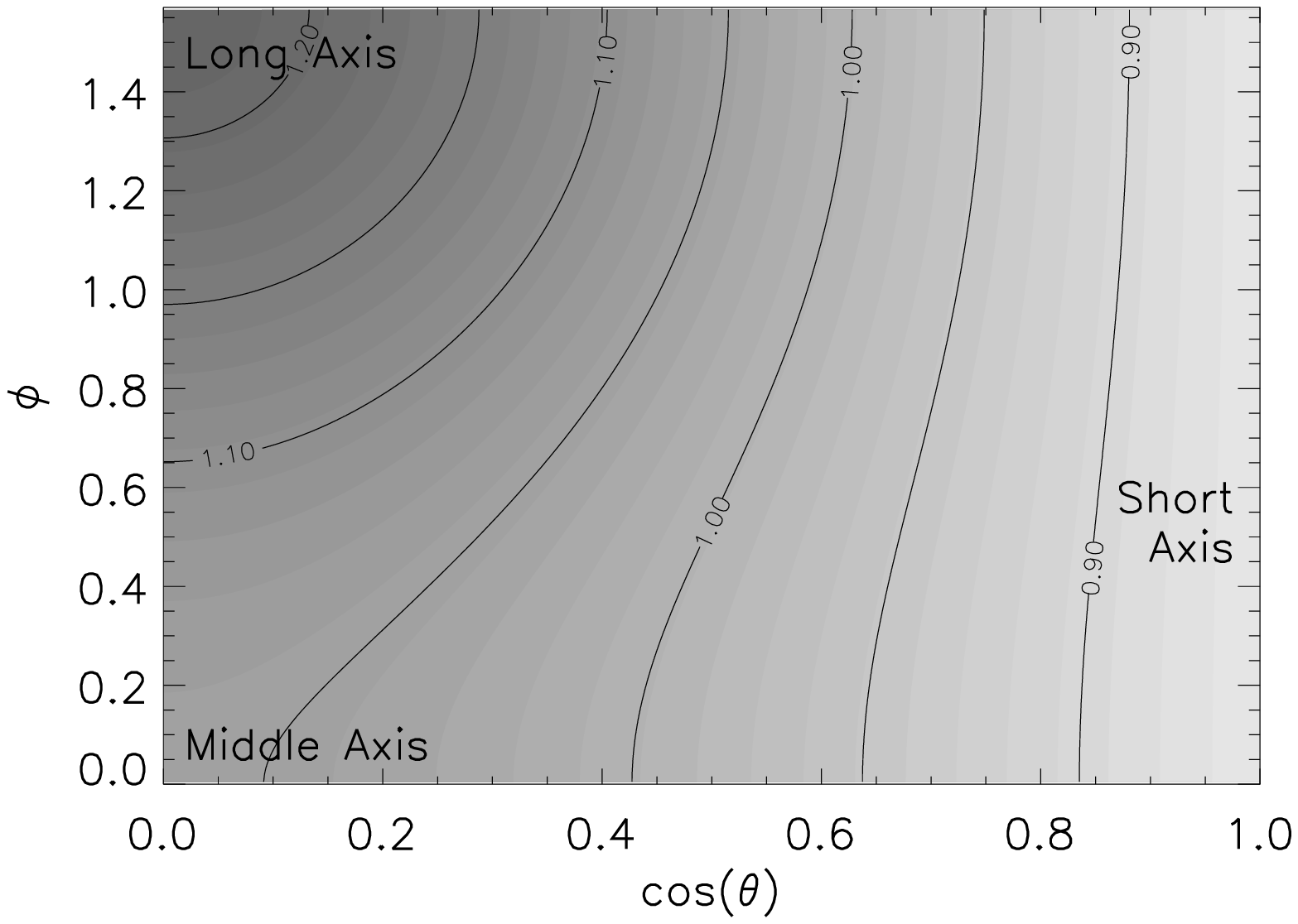}
\plotone{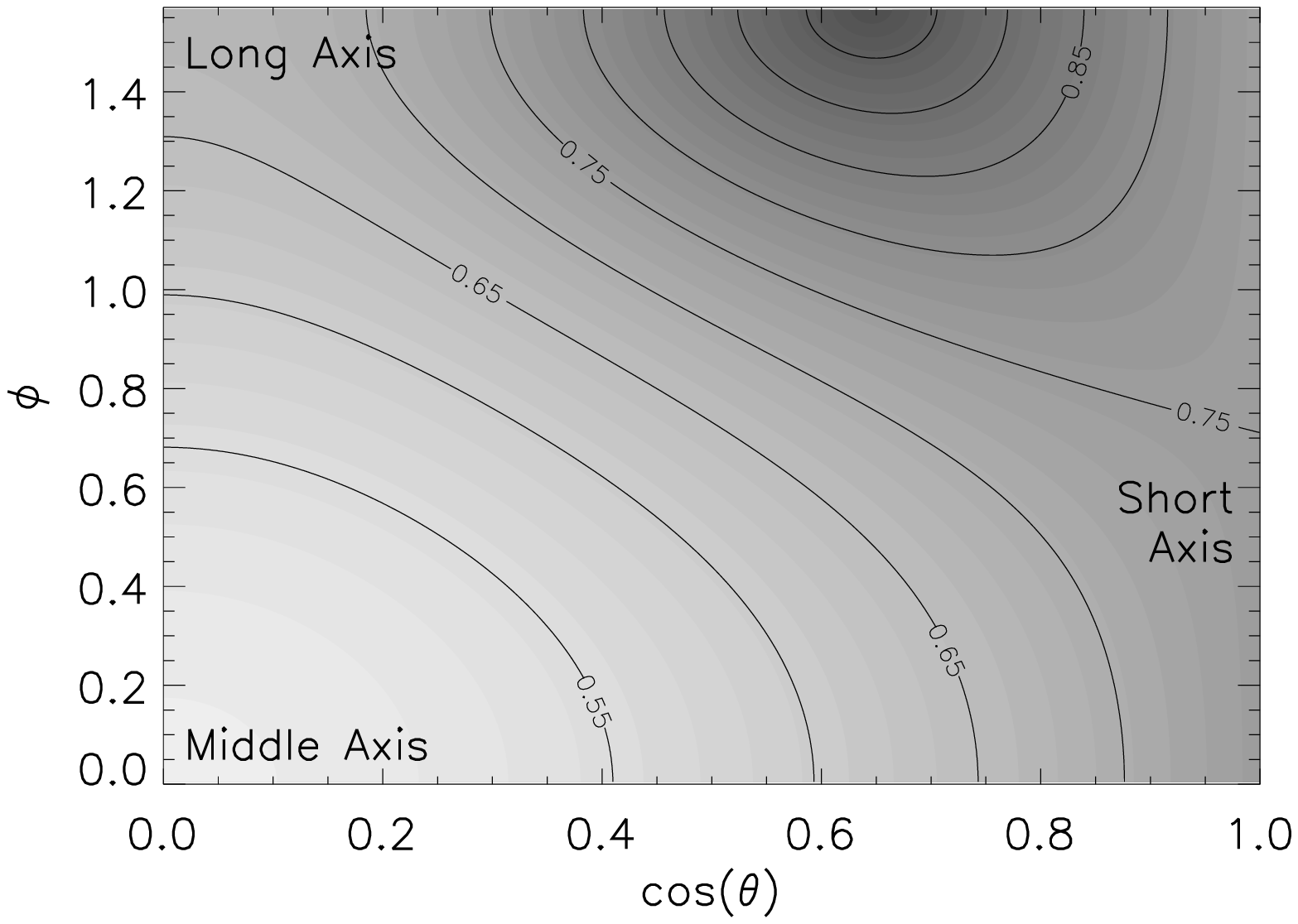}
\caption{{\it Top panel:} Einstein radius $b/b_0$ as a function of line of sight for a triaxial
isothermal profile (see Eq. \ref{eq:3dsie}) with axis ratios $q_1=0.75$ and $q_2=0.5$.
Here, $b_0$ is the Einstein radius obtained in the spherically symmetric case ($q_1=q_2=1$).
{\it Bottom panel:} Projected axis ratio for the same halo. 
Our coordinate system is such that $\cos(\theta)=1$ corresponds to projections along the
short axis of the halo, $\cos(\theta)=0, \phi=0$ corresponds to projections along the middle
axis, and $\cos(\theta)=0, \phi=\pi/2$ are projections along the long axis of the lens.  Note
projections along the long axis of the lens ($\cos(\theta)=0, \phi=\pi/2$) result in the largest
Einstein radius, whereas projections along the middle axis ($\cos(\theta)=0, \phi=0$) 
maximize the ellipticity of the projected density profile. }
\label{fig:orient}
\end{figure} 


Figure \ref{fig:orient} shows the ratio $b(\los)/b_0$ where $b_0$ is the
Einstein radius of an SIS with velocity dispersion $\sigma_v$, as well as the projected axis ratio $q(\los)$,
for an isothermal ellipsoid with axis ratios $q_1=0.75,\  q_2=0.5$.
We can see the Einstein radius of the projected profile is maximized when projecting along the long axis of the
halo, whereas the ellipticity is maximized when projecting along the middle axis of the halo, as it should be.
Note we have only considered the range $\theta\in[0,\pi/2]$ and
$\phi\in[0,\pi/2]$ rather than the full range of possible lines of sight $\theta\in[0,\pi/2]$ and $\phi\in[0,2\pi]$.  This is 
due to the symmetry of our model; all eight of the octants defined by the symmetry planes of the ellipsoids are
identical.

\begin{figure*}[t]
\epsscale{1.15}
\plottwo{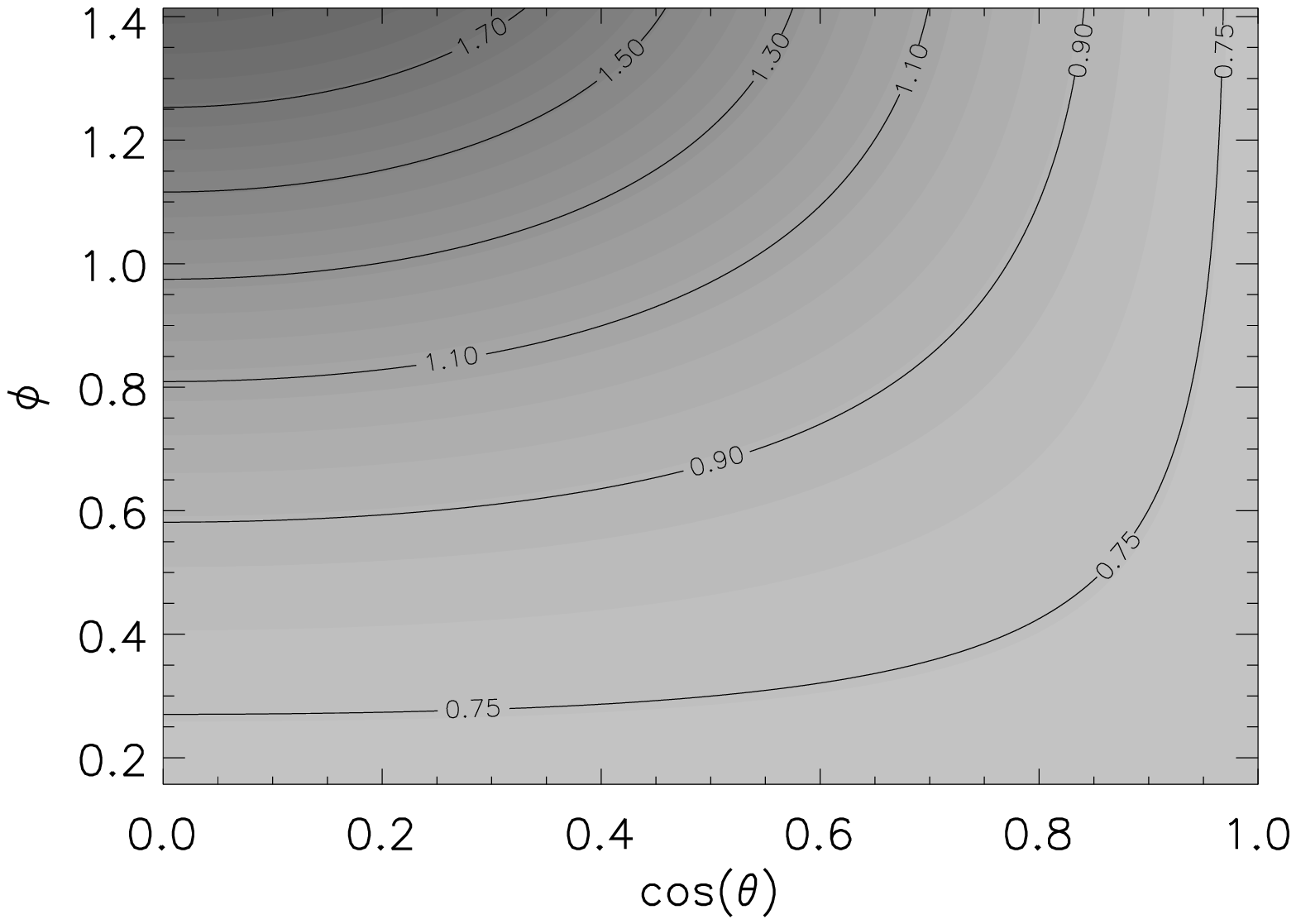}{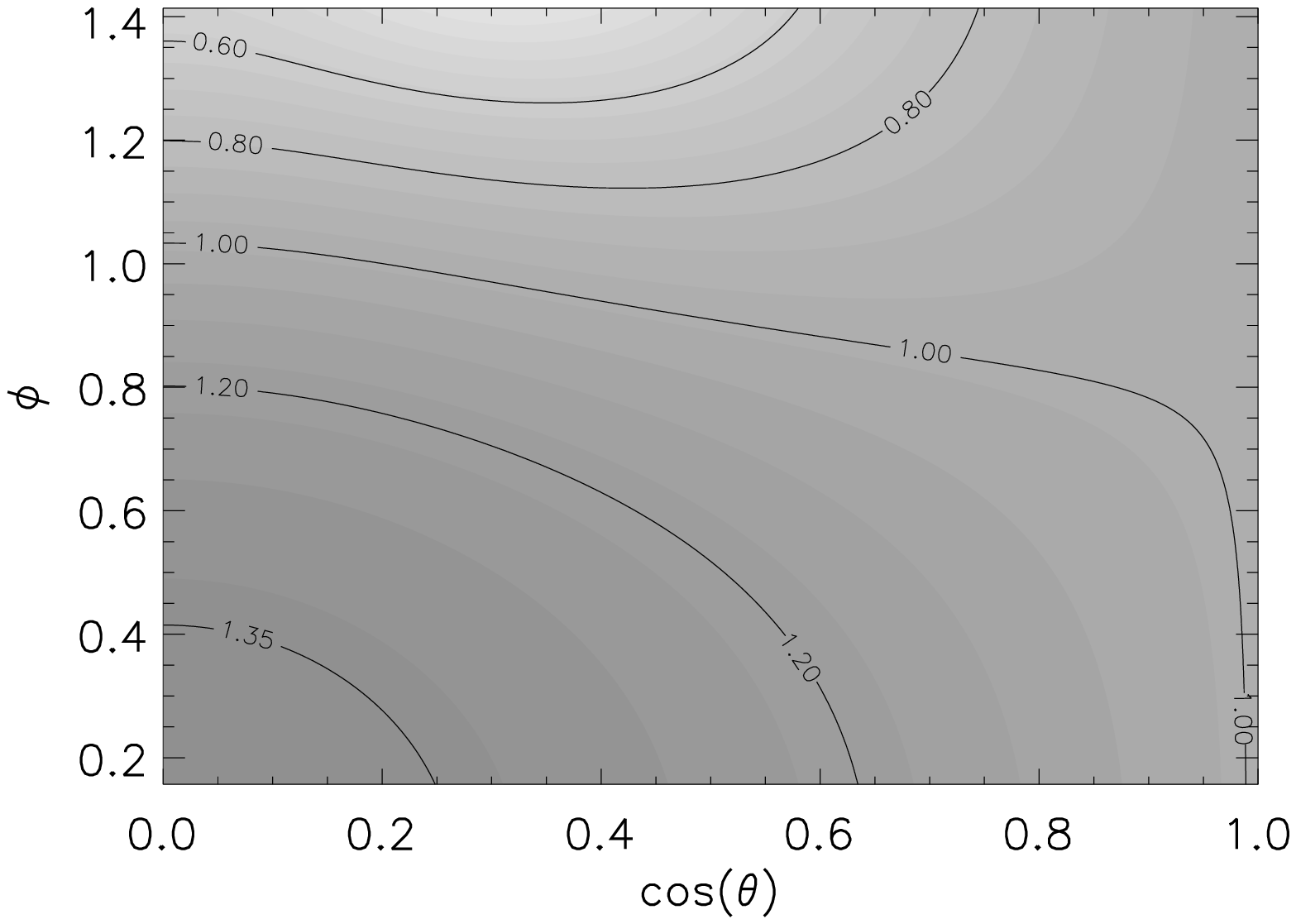}
\plottwo{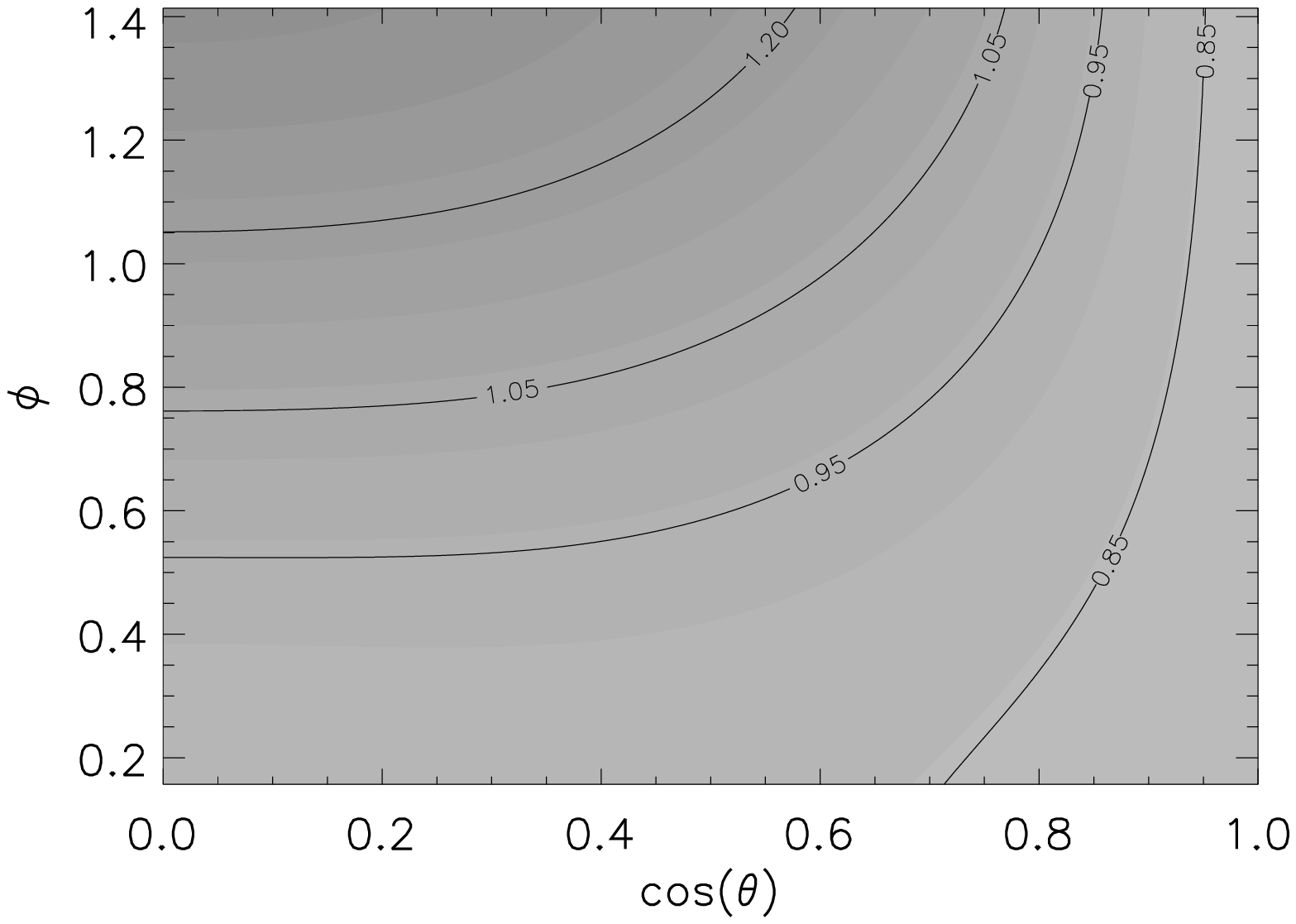}{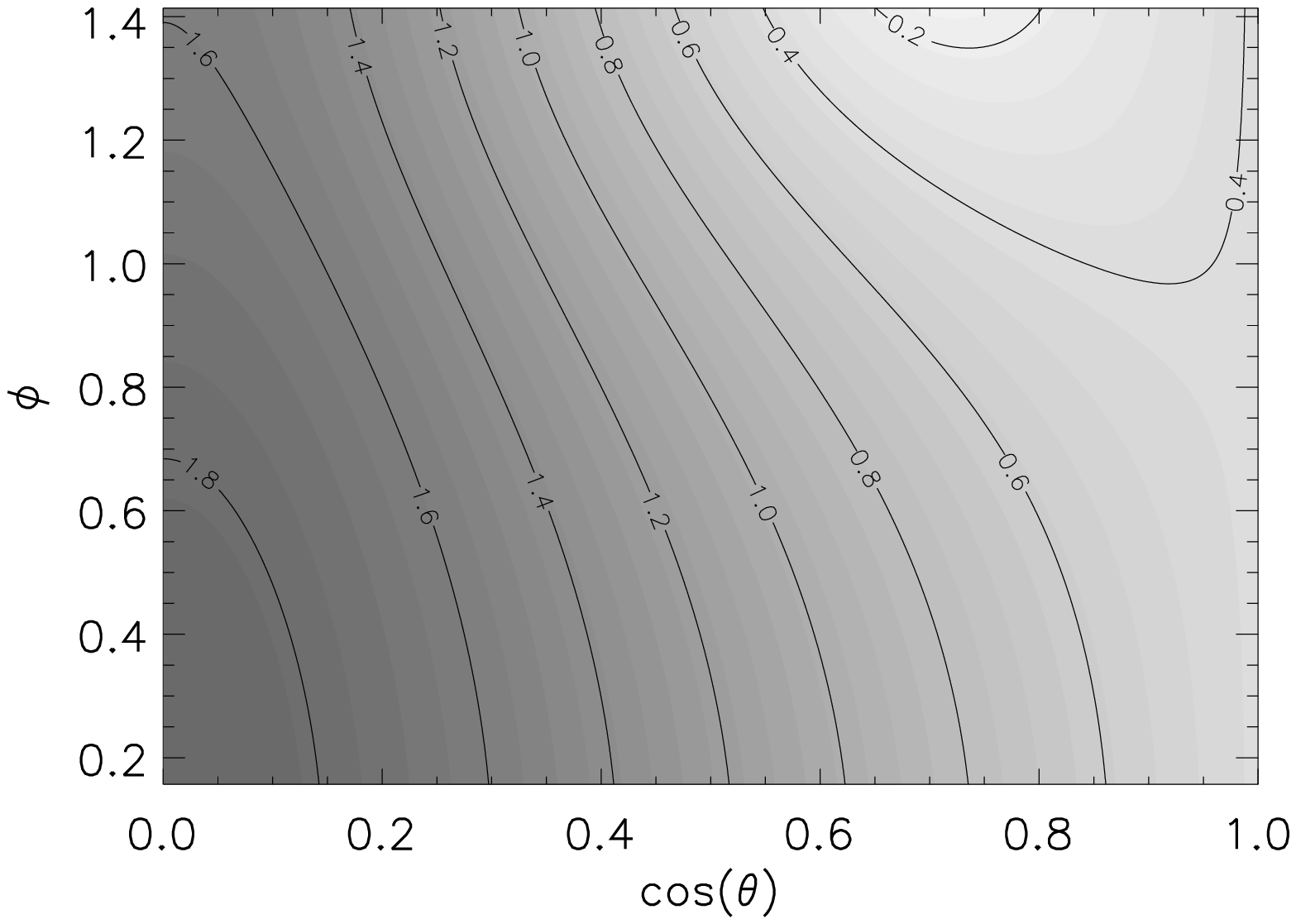}
\plottwo{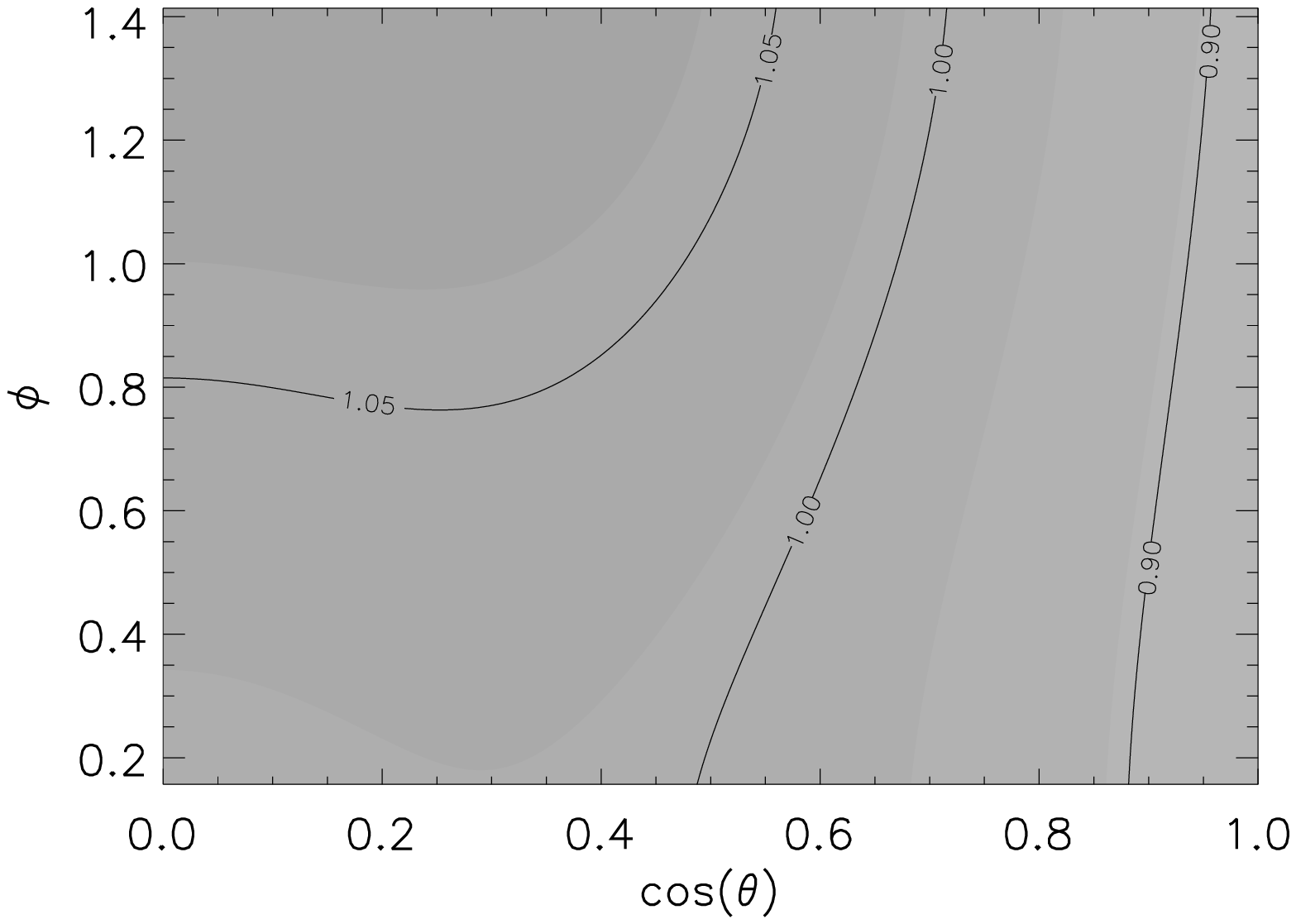}{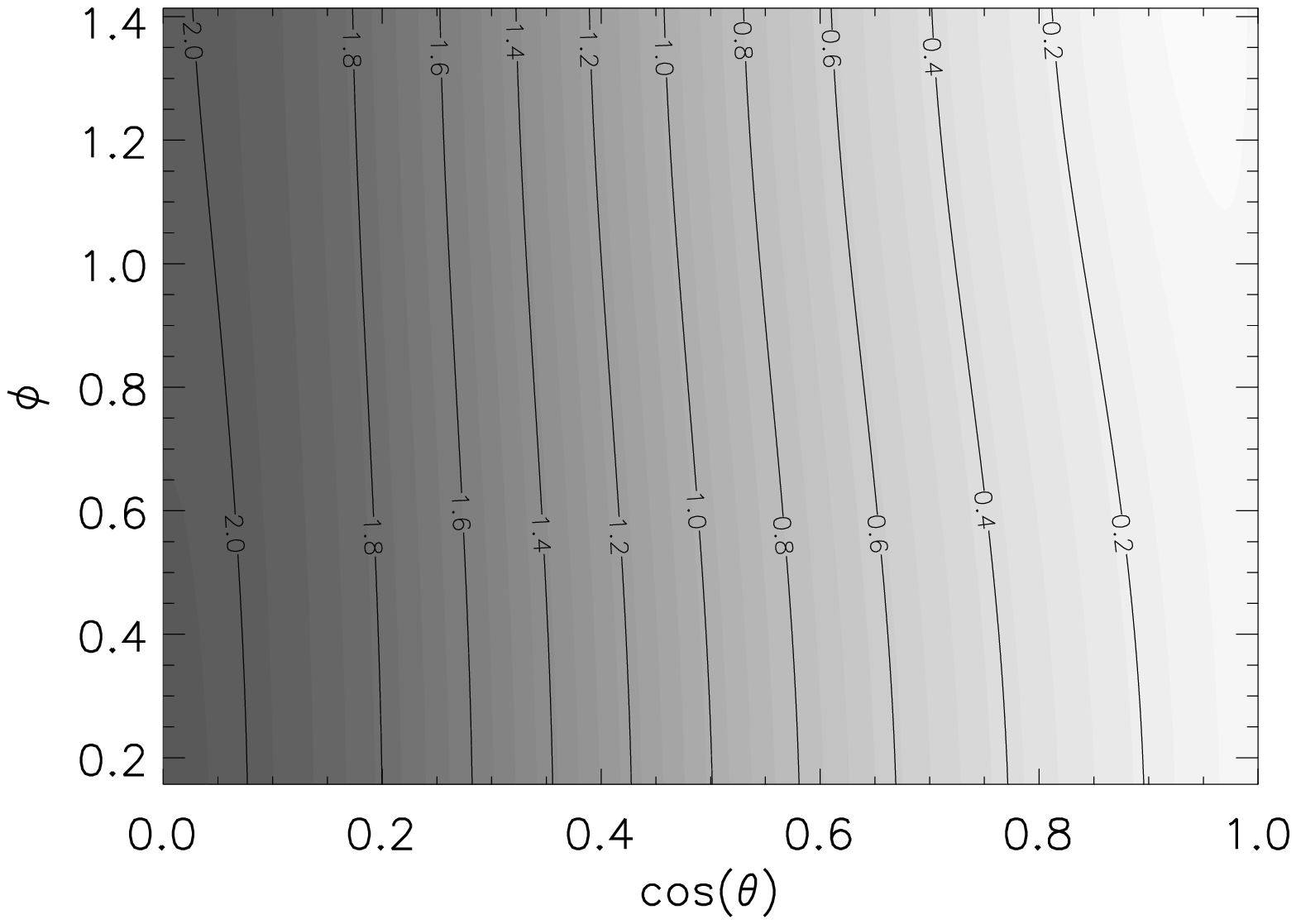}
\caption{Contours of the orientation distribution $\rho(\los)=\sigma_B/\avg{\sigma_B}$ for triaxial isothermal profiles.
The left and right columns show the distributions for double and quads respectively, while the three rows are for
three different halos: the top row is for prolate (cigar-like) halos ($T=0.9$), the middle row if a for a triaxial
halo that is neither strongly oblate nor prolate ($T=0.5$), and the bottom row is for oblate (pancake-like)
halos ($T=0.1$).  The short to long axis ratio $q_2$ was held fixed at $q_2=0.5$.  
For ease of comparison, the color scale is the same in all plots.  Note that
for doubles, the distribution of lines of sight is peaked about projections along the long axis of the lens, as we
would expect.  For quad systems, however, the distribution peaks for projections along the middle axis of the
lens, which corresponds to maximizing the ellipticity of the resulting projected profile (see Figure
\ref{fig:orient}).}
\label{fig:losdist}
\end{figure*}

Let us now go back and study the distribution of line of sights for both doubles and quads.
Figure \ref{fig:losdist} shows these distributions for three types of halos: a prolate halo, an oblate halo,
and a halo that is neither strongly oblate nor strongly prolate.  As is customary, we parameterize the halo
shape in terms of the shape parameter $T$ which is defined as
\begin{equation}
T = \frac{1-q_1^2}{1-q_2^2}.
\label{eq:shape_parameter}
\end{equation}
Note that a perfectly prolate halo ($q_1=q_2$) has $T=1$, 
whereas a perfectly oblate halo ($q_1=1$) has
$T=0$.  From top to bottom, the halo shape parameters used to produce Figure \ref{fig:losdist} are
$T=0.9$ (cigar shape), $T=0.5$ (neither strongly oblate nor strongly prolate), and $T=0.1$ 
(pancake shape). 
The axis ratio $q_2$ was held fixed at $q_2=0.5$.
Finally, the left column is the distribution of lines of sight for double systems, 
whereas the right column is the distribution for quads.  For ease of comparison, the color scale has
been kept fixed in all plots.


Let us begin by looking in detail at the doubles column first.  As is to be expected, the distribution of lines
of sight is peaked for projections along the long axis of the lens, as this line of sight maximizes the Einstein
radius of the projected profile.  Moreover, the distribution is very sharply peaked for cigar-like halos (top row),
but is rather flat for pancake-like halos (bottom row).   The reason that the distribution of lines of sight 
for pancake-like halos is so flat is simple: for an oblate halo, projecting along either the long or medium
axis of the halo results in a large Einstein radius, but also a large ellipticity, so a large part of the multiply imaged
region of the source plane actually corresponds to four image configurations, taking away from the cross
section for producing doubles.   When projecting along the short axis of the lens, the Einstein radius is
minimized, but the projected mass distribution is nearly spherical, so the majority of the multiply-imaged region
produces only doubles.

The column corresponding to quads has much more interesting structure.   First, note that {\it the distribution
of line of sights for quad lenses peaks for projections along the middle axis of the lens rather than the long
axis of the lens.}   As noted earlier, projections along the middle axis of the lens maximize the ellipticity of the
projected profile, so relative to projections along the long axis of the lens, it is evident that the increase in ellipticity 
more than offsets the slightly smaller Einstein radii for the purposes of enhancing the lensing cross section for
producing quad systems.  It is also interesting to note that while the peak of the distribution is always 
clearly about the middle axis of the lens, the shape of the distribution varies considerably in going 
from prolate halos to oblate halos.   In particular, note that for prolate halos the peak about the middle 
axis is relatively narrow.  What is more, projections along the short axis of the lens are more likely than projections
along the long axis because the latter minimizes the ellipticity of the projected profile.  
For oblate halos, on the other hand, projections along the long axis of the lens are almost
as likely as projections along the middle axis.  This is simply because for such halos, there is little
difference in the ellipticity of the projected profile between projections along the middle and long axis of the
halos.  Consequently, both axes result in highly effective quad lenses.  Note too that for pancake-like halos, 
projections along the short axis are strongly avoided, since this projection minimizes both the Einstein radius
and the projected axis ratio of the lens.  

In short, then, prolate halos and oblate halos will have very different orientation distributions: for prolate halos,
nearly all doubles will be due to projections along the long axis of the lens, while most quads will be due to 
projections along the middle axis of the lens, followed by projections along the short axis.  For oblate halos,
however, all halo orientations are almost equally likely in the case of doubly imaged systems, whereas quads
strongly avoid projections along the short axis of the halo.

The remainder of the paper will explore whether these results have a significant impact on the statistical
properties of the halo population.  Specifically, we will first consider the ellipticity distribution 
of lensing galaxies compared to that of galaxies as a whole.  We will then discuss how these 
results affect the predicted quad-to-double ratio, and finally, we will investigate whether 
lensing halos are expected to have a significantly biased projected substructure mass fraction.


\section{The Projected Axis Ratios of Lensing Halos}
\label{sec:axis}

As mentioned in the introduction, if one assumes that 
the ellipticity of the light and that
of the mass are monotonically related, then one would naively expect that 
lensing galaxies that produce quads ought to be more elliptical than the average galaxy because
the lensing cross section for quads increases with increasing ellipticity.
Similarly,
galaxies that produce doubles should tend to be more spherical.  In this section,
we discuss the impact of halo triaxiality on the distribution of axis ratios for double and quad lenses.


\begin{figure}[t]
\epsscale{1.2}
\plotone{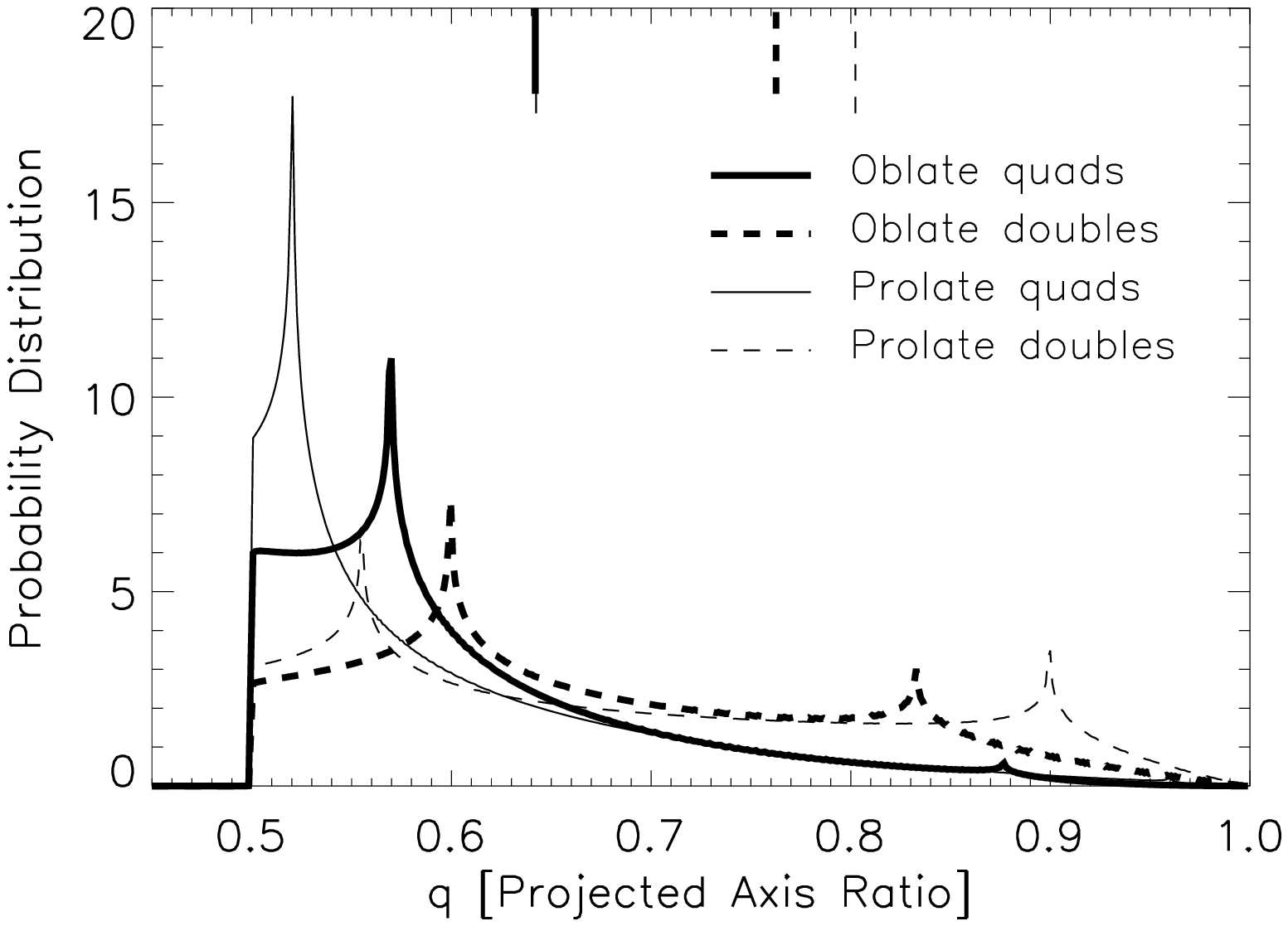}
\caption{Distribution of projected axis ratios for quads and doubles for a sample oblate halo ($T=0.1$)
and a sample prolate halo ($T=0.9$).   The distribution of projected axis ratios for doubles and quads
are fairly different, with that of quads being peaked at low axis ratios (high ellipticities) while that of doubles is
relatively flat.  The quads distribution for prolate halos is significantly more peaked than that of oblate
halos, suggesting that the distribution of projected axis ratios in quadruply imaged systems may help constrain
the shape of the three dimensional matter distribution of halos.  Also shown in the figure as lines on the top axis are
the values of $q_{0.75}$, the projected axis ratio for which $75\%$ of the lenses have $q\leq q_{0.75}$ (the oblate
and prolate values for $q_{0.75}$ are nearly identical).  Note that
the difference in $q_{0.75}$ between quads and doubles is quite pronounced, with $\Delta q_{0.75}> 0.1$
for both oblate and prolate halos.}
\label{fig:qproj_dist}
\end{figure} 


Given a line of sight $\los$, we can compute the axis ratio $q(\los)$ of the projected mass distribution (see Eq.
\ref{eq:qproj}).  Using the distribution of lines of sight $\rho(\los)$, one can then easily compute the distribution of
projected axis ratios $q$ for a sample of lenses via
\begin{equation}
\rho(q|q_1,q_2) = \int \frac{d\los}{2\pi} \rho(\los) \delta_D(q(\los|q_1,q_2)-q).
\end{equation} 
Figure \ref{fig:qproj_dist} shows the distribution of the projected axis ratio of both quad and double systems
for the sample pancake-like (oblate, $T=0.1$) and cigar-like (prolate, $T=0.9$)
halos from Figure \ref{fig:losdist}.
As is to be expected, the distribution for quad systems is considerably skewed 
towards high ellipticity systems, whereas the distribution for doubles is much flatter.  Moreover, the quads distribution
is significantly more skewed for prolate (cigar-like) systems than for oblate (pancake-like) halos.
Based on Figure \ref{fig:qproj_dist}, we have attempted to
distill the difference between quads and doubles into a single number.  We define
the axis ratio $q_{0.75}$ as the axis ratio for which $75\%$ of the lenses have axis ratios 
$q\leq q_{0.75}$.\footnote{The $75\%$ number is selected in a somewhat ad hoc manner.
Basically, we wanted $q_X$ to fall past the large prominent peak seen in Figure \ref{fig:qproj_dist},
and in that sense $X=80\%$ or $X=90\%$ would work just as well.  On the other hand, observational
estimates of $q_X$ for $X$ close to unity would be quite difficult, so to some extent we wanted $X$ 
to be as small as possible.  We chose $X=75\%$ as a reasonable value.}  The value
$q_{0.75}$ for quads and doubles for both sample halos is also shown in Figure \ref{fig:qproj_dist} as
lines along the top axis of the plot.  It is clear that the projected axis ratio $q_{0.75}$ for doubles and quads is
very different, with $\Delta q_{0.75}>0.1$ for both oblate and prolate halos.

\begin{figure}[t]
\epsscale{1.2}
\plotone{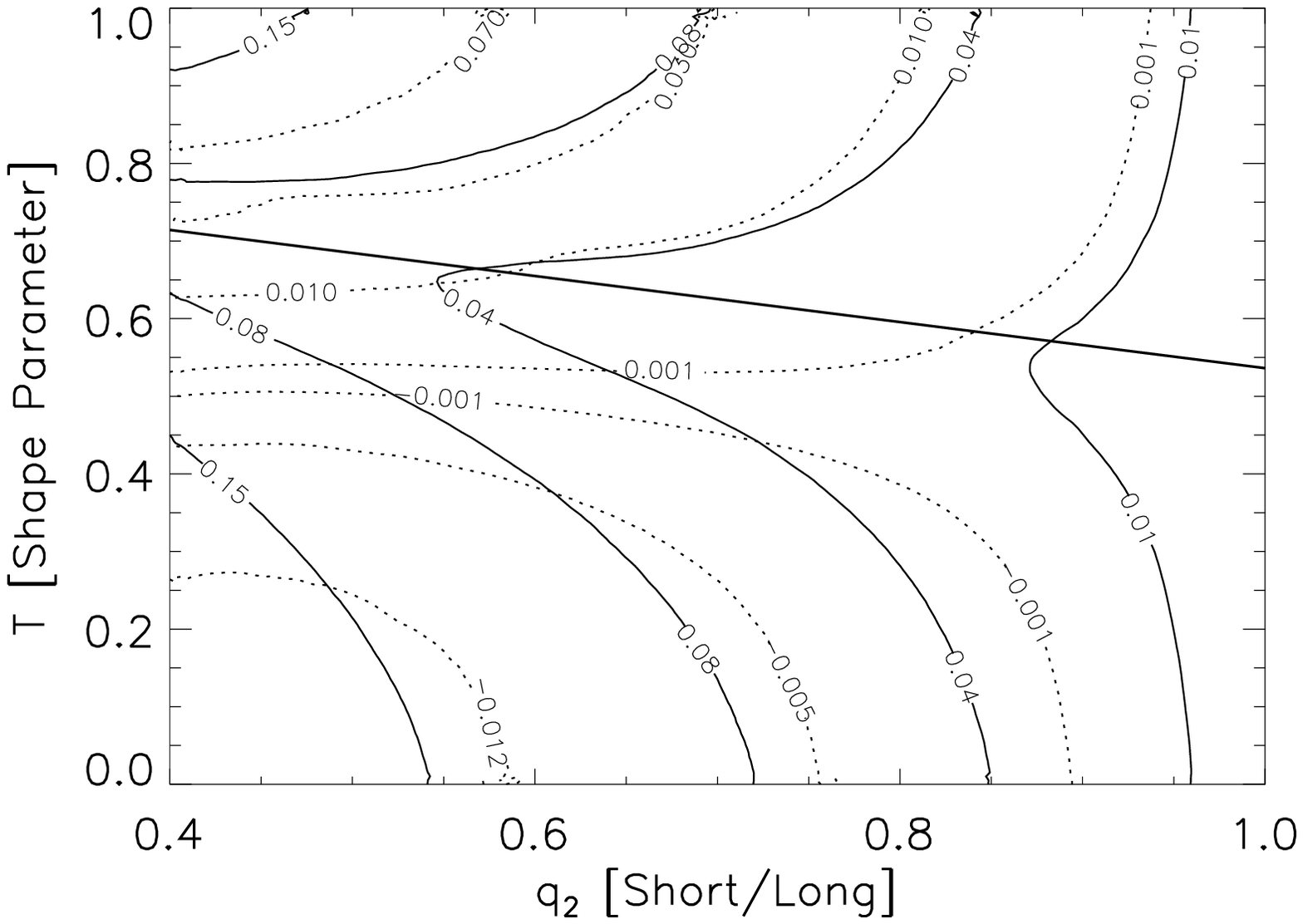}
\caption{Difference $\Delta q_{0.75}$ between doubles and quads ($q_{0.75}^{doubles}-q_{0.75}^{quads}$, 
solid line) and between doubles and
the overall halo population ($q_{0.75}^{doubles}-q_{0.75}^{halos}$, dotted lines) as a function of the 
small-to-large axis ratio $q_2$ and the shape parameter $T$
(see Figure \ref{fig:qproj_dist} for the definition of $q_{0.75}$).  For low values of $q_2$
($q_2\approx 0.5$), $\Delta q_{0.75}$ between doubles and quads is typically large, of order $0.1$.  
The minimum in $\Delta q_{0.75}$ between quads and doubles occurs for axis ratios
$q_2\approx q_1^2$, the latter relation being shown above with the thick, solid straight line. 
By comparing the axis ratio
distribution for quad lenses to those of doubles and those of the overall halo population, 
one could in principle determine if halos are typically oblate (pancake-like), prolate (cigar-like), 
or in between. }
\label{fig:axis_ratio}
\end{figure}

Figure \ref{fig:axis_ratio} shows the difference $\Delta q_{0.75}$ between doubles and quads (i.e. $q_{0.75}^{doubles}-
q_{0.75}^{quads}$, solid line) and
between doubles and the overall halo populations (i.e. $q_{0.75}^{doubles}-q_{0.75}^{halos}$, dotted line) as 
a function of  the axis ratios $q_1$ and $q_2$.   However, rather than using $q_1$ as an axis,  we follow
standard practice and parameterize the shape of the halo in terms of the shape parameter $T$ defined in
Eq. \ref{eq:shape_parameter}.
There are several interesting things to be gathered from 
Figure \ref{fig:axis_ratio}.
First, when comparing doubles to quads, note that while $\Delta q_{0.75}$ is indeed large ($q_{0.75}\gtrsim 0.1$) 
for both prolate
and oblate halos, the difference can be larger for oblate halos than for prolate halos.  Moreover, 
note that 
in going from oblate to prolate halos, the difference $\Delta q_{0.75}$ goes through a minimum when
$q_2 \approx q_1^2$ (solid line), in which case values 
as low as $\Delta q_{0.75} \approx 0.05$ for $q_2\approx 0.5$ are possible.  
Turning now to the comparison between
doubles and random halos, we see that the difference in $q_{0.75}$ for these two halo populations becomes
negligible in the case of oblate halos, reflecting the near uniform distribution of lines of sights for doubles
for oblate halos (see Figure \ref{fig:losdist}). On the other hand, the fact that most prolate doubles are seen
along the long axis of the halo implies that $\Delta q_{0.75}$ between doubles and random halos
must be significant, and thus doubles tend to be more circular than the typical halo.


In short, then, the quantity $\Delta q_{0.75}$ between doubles and quads and between doubles and random halos
can, at least in principle, help determine whether most halos are oblate or prolate.  If halos are prolate, the difference 
$\Delta q_{0.75}$ between doubles and random halos is large.  If this difference is small, we can then look at the 
difference $\Delta q_{0.75}$ between doubles and quads.  If this last difference is large, then halos are typically oblate, 
whereas if the difference is small, then halos are neither strongly oblate nor strongly prolate and $q_2\approx q_1^2$.

In practice, however, the above test is difficult to execute.  In particular, while lens modeling can provide some
measure of the axis ratio $q$ in quad systems, there remains a fair amount of uncertainty due to the approximate
degeneracy between galaxy ellipticity and external shear \citep[see e.g.][]{keetonetal97}.  
This degeneracy is even stronger for doubly-imaged
systems, and worse, there is no way of determining the axis ratio of the mass for non-lensing galaxies.  Fortunately,
at the scales relevant for strong lensing ($\lesssim 5\ \kpc$), baryons dominate the total matter budget in early type
galaxies \citep[][]{rusinetal03}, so one expects that the dark matter distribution in these systems
will have the same ellipticity and orientation as the baryons.   
Observationally, \citet[][]{keetonetal98} \citep[see also][]{keetonetal97}
compared the projected ellipticity of the light in lensing galaxies to the ellipticity recovered from explicit
lens modeling, and found that the light and the mass tend to be very closely aligned, though the
magnitude of the ellipticities is not clearly correlated and the modest quality of the photometry available
at the time made their ellipticity measurements difficult.
Moreover, the galaxy sample \citet[][]{keetonetal98} included many galaxies that had non-negligible
environments that were not incorporated into the model.  More
recently, a detailed study of the Sloan Lens ACS Survey \citep[SLACS][]{boltonetal06} with more isolated galaxies
supports the hypothesis that the ellipticity of the light is in fact extremely well matched to the ellipticity of the projected 
mass, at least on scales comparable to the Einstein radii of the galaxies \citep[][]{koopmansetal06}.\footnote{We note,
however, that SLAC lenses tend to have Einstein radii that are quite comparable to their optical radii, so the agreement
is really expected.  In principle, a discrepancy could exist for lenses with larger Einstein radii for which the total mass
has a larger dark matter component.}
Thus, for the purposes of this
work, we simply take the isophotal axis ratio of lensing galaxies to be identical to the total matter 
axis ratio for the purposes of investigating whether lens biasing can be detected in current lensing samples.

\begin{figure}[t]
\epsscale{1.2}
\plotone{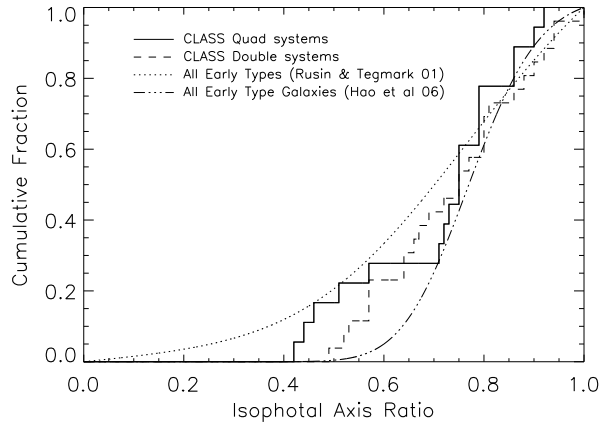}
\caption{Cumulative distribution of the isophotal axis ratio for early type galaxies.  The step-lines
shown correspond to galaxies that produce quads (solid) and doubles (dashed).  The smooth curves
show the overall
early type galaxy population in the Coma clusters \citep[dotted, see][]{rusintegmark01} and
in the SDSS DR4 \citep[dashed, see][]{haoetal06}.  The axis
ratio distribution for lensing galaxies is consistent with both model distributions, and with each
other, suggesting that the typical halo is neither
strongly oblate nor prolate, with axis ratios $q_1$ and $q_2$ such that $q_2\approx q_1^2$.}
\label{fig:cumq}
\end{figure}

Figure \ref{fig:cumq} shows the cumulative
distribution of isophotal axis ratios for quad lenses (solid) and double lenses (dashed) for all lensing galaxies
in the CASTLES\footnote{http://cfa-www.harvard.edu/castles/} database 
with isophotal axis ratios measurements.\footnote{This data was kindly provided 
by Emilio Falco, private communication.}    Of course, the selection function for this sample is impossible to
quantify objectively, but our intent is simply to see whether any differences between lensing galaxies and
random galaxies can be found.
Also shown in the figure are the axis ratio distributions of early type galaxies as reported by 
two different groups:
the dotted line shown is the fit used by \citet[][]{rusintegmark01} to model the distribution
of axis ratios in early type galaxies based on measurements by \citet[][]{jorgensenfranx94}, and 
is also quite close to the distribution recovered by \citet[][]{lambasetal92}.  The
dashed-dotted line is the axis ratio distribution obtained by \citet[][]{haoetal06} using the SDSS Data Release 4
photometric catalog, and is a very close match to the distribution recovered by \citet[][]{fasanovio91}.   
\citet[][]{haoetal06} noted that it is unclear why these two distributions differ, though \citet[][]{keetonetal97}
note that such a difference can easily arise depending on whether S0 galaxies are included in the
galaxy sample or not (with S0 galaxies being more elliptical).  Here, we simply consider both distributions.
 
Given that the axis ratio distribution for both quads and doubles largely fall in between the two model 
distributions we considered, it is immediately obvious that no robust results can be obtained at this
time.  Specifically, uncertainties in the details of the selection function of the galaxies used to construct
the isophotal axis ratios are a significant systematic.  More formally,
using a KS-test, we find that the isophotal axis ratio distributions of both quad and double lens galaxies
are consistent with that of the early type galaxy population as a whole (irrespective of which model distribution
we choose) and with each other as well.    Interestingly, whether or not
we restrict ourselves to galaxies that are isolated or whether we include all lensing galaxies
does not appear to change the result in any way.
Naively, then, the consistency of the axis ratio distributions
suggests that halos are typically neither strongly oblate nor prolate, but rather somewhere in between,
where the quantity $\Delta q_{0.75}$ exhibits a minimum, which occurs at $q_2\approx q_1^2$.


Given that current lens samples are too small for detecting any difference on the ellipticities of quadruply and 
doubly imaged systems, it is worth asking whether or not a detection is possible in principle.  That is, 
how many lenses must one have in order to detect quad systems as being more elliptical than doubles?
To answer this question, we need to first assume a simple model for the distribution of axis ratios $Q(q_1,q_2)$,
with which one could then compute the resulting projected axis ratio distributions for doubles, quads, and the
galaxy population at large.   We should note here, however, that in detail our results will depend on the
adopted distribution $Q(q_1,q_2)$, which is not known.

It is not immediately obvious what the most correct model distribution $Q(q_1,q_2)$ should be.  
While there have been many studies that have investigated the distribution of axis ratios
of dark matter halos in simulations \citep[see e.g.][]{warrenetal92,jingsuto02,bailinsteinmetz05},  
it has become clear
that the distribution itself depends on many variables, including halo mass \citep[][]{kasunevrard05,bettetal07},
radius at which the shape of the halo is measured \citep[][]{hayashietal07}, halo environment \citep[][]{hahnetal07},
and whether the halo under
consideration is a parent halo or a subhalo of a larger object \citep[][]{kuhlenetal07}.  Adding to these difficulties
is the fact that different authors use different definitions and methods for measuring the shapes of halos,
which forces one to go to great lengths in order to ensure a fair comparison of the results from different groups
\citep[see for example][]{allgoodetal06}.  Even more problematic that all of these difficulties, however, is
the fact that not only can the distributions of baryons have a different shape from the dark 
matter \citep[][]{gottloberyepes07}, baryons dominate the mass budget in the halo regions where strong 
lensing occurs, and can
therefore dramatically impact halo shapes at those scales \citep[][]{kazantzidisetal04,bailinetal05,gustafssonetal06}.  
Since our intent here is simply
to provide a rough estimate of the number of lenses required to detect a significant difference in the ellipticities 
of quad and double systems, we simply adopt a fiducial model that is based primarily on the results 
of \citet[][]{allgoodetal06} and \citet[][]{kazantzidisetal04}, and use it to estimate the number of lenses necessary
to detect the larger ellipticity of quad systems.    
Specifically, \citet[][]{allgoodetal06} obtain that for an $M_*$ halo the distribution 
of the short-to-long axis ratio of dark matter halos is Gaussian with a mean of $\avg{q_2} = 0.54$ and a standard
deviation $\sigma_{q_2}=0.1$.   As noted by \citet[][]{kazantzidisetal04}, baryonic cooling tends to 
circularize the mass profiles of halos, so we adopt instead a somewhat larger ratio $\avg{q_2}=0.65$, but retain 
the dispersion $\sigma_{q_2}=0.1$.  The adopted value for $\avg{q_2}$ is larger than that obtained
from dissipationless simulations, but smaller than that found in the simulations of
\citet[][]{kazantzidisetal04}, as the latter suffer from the well known over-cooling problem and therefore
overestimate the impact of baryons on the profiles.  In addition, 
we truncate the distribution at $q_2=0.4$, as the expressions for 
the lensing cross sections are no longer valid for systems with projected axis ratios 
below $0.4$.\footnote{For SIE profiles, if the projected axis ratio $q<0.4$, then naked cusp
configuration appear.  Since the analytical formulae we used to compute $\sigma(q)$
all compute the area contained within the tangential caustic, it follows that for $q<0.4$, 
our cross section estimates would correspond to the total cross section for producing
either quads or naked cusps.  To avoid this complication, we simply truncate our
axis ratio distribution at $q_2=0.4$.  Note however that since $q_2=0.4$
is already $2.5\sigma$ away from the adopted mean we expect the
introduced cutoff to have a negligible impact on our results.}
Finally, the value of the intermediate axis $q_1$ is obtain following the 
model of \citet[][]{allgoodetal06}
\citep[itself based on the work by][]{jingsuto02}, namely, the quantity $p=q_2/q_1$ is drawn from the distribution
\begin{equation}
\rho(p|s) = \frac{3}{2(1-s)}\left[1 - \left(\frac{2p-1-s}{1-s}\right)^2\right]
\end{equation}
where $s=\mbox{min}(0.55,q_2)$.


\begin{figure}[t]
\epsscale{1.2}
\plotone{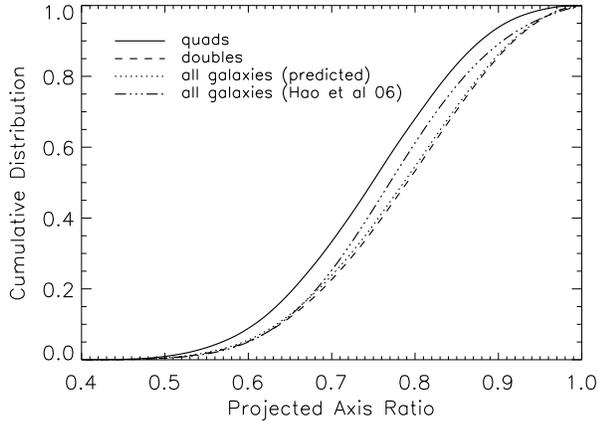}
\caption{Cumulative distribution of the isophotal projected axis ratio for all galaxies, quad lenses,
and double lenses, using the fiducial model discussed in the text.  For comparison, we also show the
observed distribution of isophotal axis ratios in the SDSS DR4 as measured by \citet[][]{haoetal06}. 
Keep in mind that the model is not meant to be a fit to the data from \citet[][]{haoetal06}, but rather is
only meant as a useful approximate model that allows us to estimate the number of lenses necessary
to distinguish between the ellipticity distribution of quads and doubles. 
We find that $N\approx 350$ lenses are needed to empirically distinguish the two distributions at
the $95\%$ confidence level.}
\label{fig:predict_qproj}
\end{figure} 


Figure \ref{fig:predict_qproj} shows the cumulative distributions of the predicted isophotal axis ratios 
for all galaxies, as well as for quad and double systems.  Also shown for reference are the axis ratio 
measurements of early type galaxies by \citet[][]{haoetal06} using SDSS DR4 data.  Note that, as we 
expected, the difference in the axis ratio $q_{0.75}$ between doubles and quads is of order $0.05$.
The maximum vertical distance between the cumulative distributions functions for quads and doubles
is $D\approx 0.15$, which, using a KS-statistic, implies that roughly $300$ lenses 
($150$ quads,  $150$ doubles) with good isophotal measurements are necessary to detect the difference
between the two distributions at the  $95\%$ confidence level.  A $5\sigma$ detection would 
require $\approx 1,400$ lenses.  Such large number of lenses is larger
than the current list of known lensing systems, but is certainly within the realm of what one may
expect from future lens searches \citep[see e.g.][]{koopmansetal04,marshalletal05}.


\section{Triaxiality and Predictions for the Quad-to-Double Ratio}
\label{sec:ratio}

We showed above that halo triaxiality can have an important impact on the distribution of axis ratios for
lensing galaxies.  Since the projected axis ratio of a halo plays a key role in the expected quad-to-double ratio
of lensing galaxies, it is easy to see that triaxiality should also affect this statistical observable.  This
is the problem we wish to consider now: how does triaxiality affect the quad-to-double ratio of lensing
galaxies?

Consider first equation \ref{eq:numlens}.  For our semi-analytic case, the halo parameters $\bm{P}$ that
determine the mass distribution of the halo
are simply the halo velocity dispersion and its two axis ratios $q_1$ and $q_2$.   What is more, we
saw that if we define $b_0(\sigma_v)$ as the Einstein radius of an SIS of velocity dispersion $\sigma_v$, then 
the ratio $\sigma/b_0^2$ depends only on the axis ratios $q_1$ and $q_2$.  If we make the further assumption 
that the distribution of halo parameters is separable, i.e. that 
\begin{equation}
\frac{dn_{halos}}{dz_hd\sigma_vdq_1dq_2} = \frac{dn_{halos}}{dz_hd\sigma_v} Q(q_1,q_2),
\end{equation}
then it is easy to see that the \it ratio \rm of the total number of quad systems to double systems depends
only on the distribution of axis ratios $Q(q_1,q_2)$ because the overall scaling of the lensing cross 
sections for both doubles and quads just factors out of the problem.  Thus, the ratio of quad-to-doubles 
is given simply by
\begin{equation}
r(q_1,q_2) = \frac{\mbox{No. of quads}}{\mbox{No. of doubles}} = 
	\frac{ \avg{\sigma_B^{(4)}|q_1,q_2} }{ \avg{\sigma_B^{(2)}| q_1,q_2} }.
\end{equation}


\begin{figure}[t]
\epsscale{1.2}
\plotone{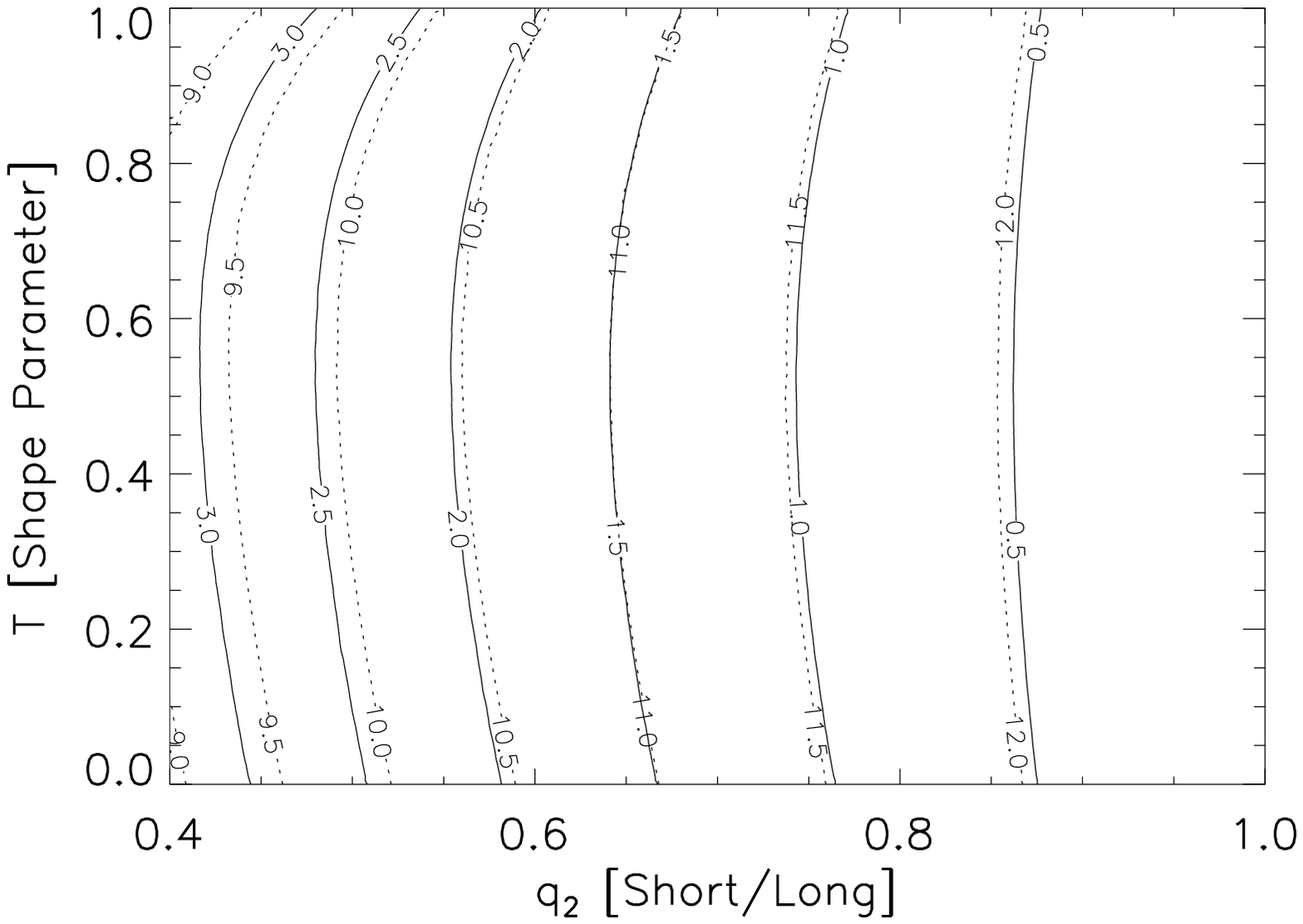}
\plotone{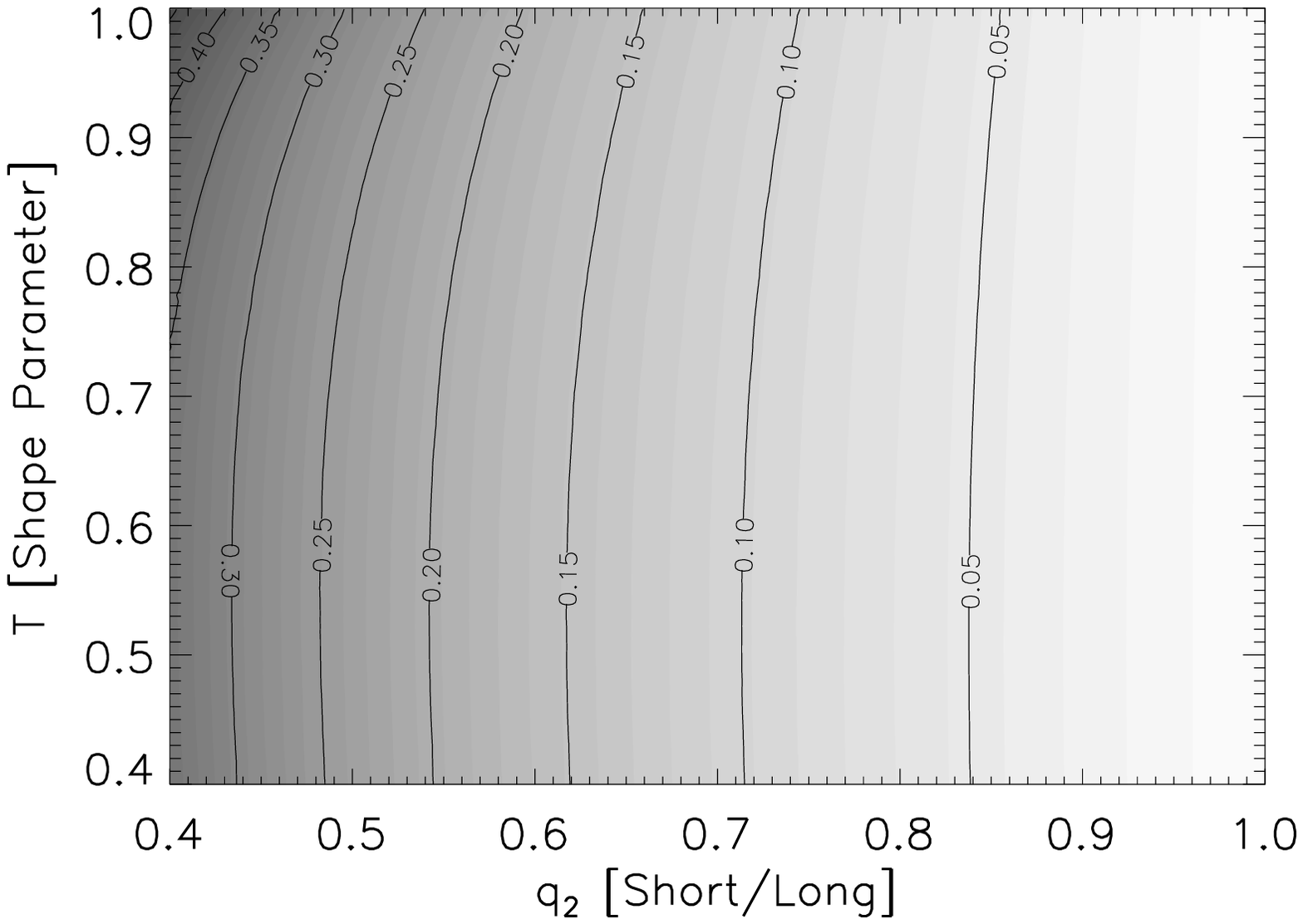}
\caption{{\it Top panel:} The dimensionless biased lensing cross section $\sigma_B/b_0^2$ for quads 
(solid) and doubles (dotted) as a function of the axis ratios $q_1$ and $q_2$ of the halo population.   
{\it Bottom panel:}  quad-to-double ratio as a function of the axis ratios $q_1$ and $q_2$ for triaxial
isothermal profiles.  We only consider the range $q_2\geq 0.4$ because below this value our
semi-analytical calculations based on \citet[][]{kormannetal94} break down.  Note the observed ratio 
of about 0.4 can only be obtained for halos that deviate strongly from spherical symmetry.  Interestingly,
all contours in both panels are very nearly horizontal, so whether halos are prolate (cigar-like) or 
oblate (pancake-like) has almost no impact on the lensing optical depths or the quad-to-double
ratio.}
\label{fig:ratio}
\end{figure} 


The top panel of Figure \ref{fig:ratio} shows the dimensionless mean biased lensing cross section 
$\avg{\sigma_B}/b_0^2$  for both doubles and quads averaged over lines of sight for a population of randomly 
oriented halos.  Also shown in the bottom panel
is the quad-to-double ratio.  As expected, large ($\gtrsim 0.3$) quad-to-doubles ratios require strong deviations 
from spherical symmetry, so $q_2$ needs to be small.  Interestingly, however, all of the contours in both the 
top and bottom panel of Figure \ref{fig:ratio} are nearly vertical: lensing cross sections are nearly independent
of halo shape.
We can understand this qualitatively as follows.   In the case of doubles, there is a tradeoff between two 
competing effects: for $1\gtrsim q_1\gg q_2$, there are many lines of sight that enhance the Einstein radius of 
the lens, but only moderately so.  For $1\gg q_1\gtrsim q_2$ on the other hand, there are only a few lines of 
sight that enhance the Einstein radius of the lens (i.e. projections along the long axis of the halo), but 
the enhancement is much greater.  Thus, the overall boost to the Einstein radius is offset by the reduced ``volume''
of lines of sight available for forming doubles and vice versa.
A similar effect occurs for quads: oblate halos make effective lenses when projected 
along either the long or middle axis of the lens, but strongly avoid the short axis, so the ``volume'' of lines of sight 
available to oblate halos is small.  Prolate halos, on the other hand, are not quite as effective as oblate halos
at making quads, but can produce quads over a larger range of possible lines of sight.

At any rate, one thing that is clear from Figure \ref{fig:ratio} is that halo shape does not have a significant impact on the
expected quad-to-double ratio.  One extremely interesting consequence of this results is that it implies that 
halo triaxiality can be properly incorporated into lensing statistics studies without greatly increasing the number
of degrees of freedom in the problem.  More explicitly, traditional lens statistics studies use as input the observed
two dimensional ellipticity distribution of early type galaxies, and approximate the effects of triaxiality by multiplying 
the usual isothermal ellipsoidal profiles with a normalization factor computed assuming halos are either all perfectly
oblate, or perfectly prolate \citep[see e.g.][]{chae03,chae07,oguri07}.   The main reason this is done, rather than
considering triaxial halos and averaging over lines of sight, is that in order to do the latter calculation, one needs
to know something about the distribution of axis ratios. We have shown, however, that such a calculation would
in fact be nearly independent of assumptions made about the intermediate axis $q_1$.
In other words, a proper calculation that weights lines of sight according to their biased lensing
cross section rather than uniform weighting (as implicitly done when taking the ellipticity distribution to be that
of early type galaxies as a whole) effectively involves no more freedom than the usual approach, the 
main difference being that the assumptions made will involve not the ellipticity distribution, but rather
the distribution of the short-to-long axis ratio $q_2$, which can itself be constrained using the
projected ellipticity distribution \citep[e.g.][]{lambasetal92}.  


\section{The Substructure Mass Fraction in the Inner Regions of Lensing Halos}
\label{sec:subs}

One of the important predictions of the CDM paradigm of structure formation is that galactic halos
contain a large amount of bound substructure within 
them \citep[see e.g.][]{whiterees78,blumenthaletal84}.   Observationally, however, both our
own galaxy and M31 have an order of magnitude less luminous companions than is predicted
if one assumes substructures have a fixed mass to light ratio \citep[][]{kauffmannetal93, klypinetal99, mooreetal99}.  
Currently, the
favored explanation for this discrepancy is that the mass to light ratio of such small structures
depends strongly on the history of the objects, and therefore only a select subset of the substructures
within the halo become luminous \citep[e.g.][]{somervilleprimack99,bensonetal02,kravtsovetal04,salesetal07}.  
While such scenarios appear to be in good agreement
with the data, it would still be desirable to provide as direct detection as possible of the remaining
dark substructures.

Motivated by the fact that dark substructures can only be discovered via their gravitational signal, 
\citet[][]{dalalkochanek02a} investigated whether the well known flux anomalies problem could be
explained as the action of dark substructures embedded within the halo of the lensing galaxy.
Using a relatively simple model, they found that in order to explain the observed flux anomalies, 
one requires a projected substructure mass fraction $f_s$ in the range $7\%>f_s>0.6\%$ at the
$90\%$ confidence level.  It was then argued by \citet[][]{maoetal04} that such a substructure mass
fraction was slightly larger than the mass fraction obtained from simulations $f_s\approx 0.5\%.$


\begin{figure}[t]
\epsscale{1.2}
\plotone{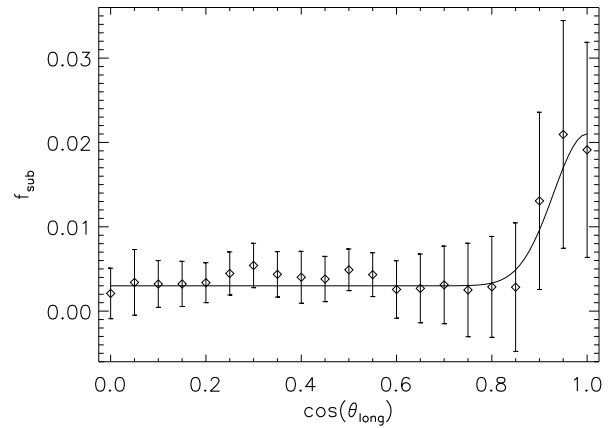}
\caption{The projected substructure mass fraction within $3\%$ of the virial radius as a function of 
$\cos(\theta_{long})$ where $\theta_{long}$ is the angle between the projection axis and the long
axis of the halo.  Error bars show the dispersion in the sample rather than the error on the mean.
The solid line shows a simple constant plus gaussian fit of the form
$f_s(x) = 0.003+0.018\exp(-(1-x)^2/0.1^2)$ where $x=\cos(\theta_{long})$.  Note projections along
the long axis of the lens have significantly higher substructure mass fractions than the typical
projection.}
\label{fig:fsublos}
\end{figure} 


Recently, it has become clear that the distribution of substructures in dark matter halos is not spherically 
symmetric, but is instead triaxial, and aligned with the major axis of the halo.  Since lensing halos
are not randomly oriented in space, the mean projected substructure mass fraction for all halos
- the $f_s\approx 0.5\%$ value obtained by \citet[][]{maoetal04} - need not be the same as the mean
substructure mass fraction for lensing halos, which would in turn affect theoretical predictions 
\citep[e.g.][]{rozoetal06,chenetal07}. Here, we use the results on substructure alignments in
numerical simulations to estimate the dependence of the projected substructure mass fraction $f_s$
on the projection axis.  More specifically, {\it assuming that substructures do
not significantly alter the biased lensing cross sections for the halos}, we compute the mean substructure mass
fraction for doubles and quad lenses as a function of the axis ratios $q_1$ and $q_2$ of the lensing halos.

We begin by presenting the substructure mass fraction $f_s$, as a function of line-of-sight in simulated
dark matter halos.  In Figure \ref{fig:fsublos}, we reproduce the distribution as presented in \citet[][]{zentner06}.  This
figure shows the mass fraction projected within $3\%$ of the virial radius as a function of the projection angle 
$\cos(\theta_{long})$  for a sample of
halos in a dissipationless $N$-body simulation of structure growth.  The angle $\theta_{long}$ is defined as the
relative angle between the projection axis and the long axis of the halo.  The data for the figure come from 26 host dark 
matter halos with masses in the range $10^{12} h^{-1}\msun < M < 10^{13} h^{-1}\msun$, and the error bar shown
represents the 
dispersion in the sample rather than the error on the mean.  The halos were drawn from a high-resolution flat, $\LCDM$
simulation with $\Omega_m=0.3,$ $\sigma_8=0.9,$ $\Omega_b h^2=0.023,$ and $h=0.7.$  Details on the simulations
can be found in \citet[][]{zentner06} or in \citet[][]{gottloberturchaninov06}.

\begin{figure}[t]
\epsscale{1.2}
\plotone{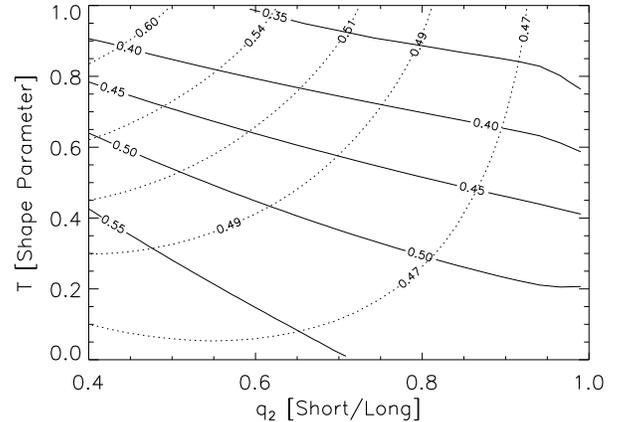}
\caption{Mean projected substructure mass fraction $\avg{f_s|q_1,q_2}$ as a function
of halo axis ratios for doubles (dotted lines) and quads (solid lines), shown 
above as percentages (i.e. we show
$100\avg{f_s|q_1,q_2}$).  The projected substructure mass fraction as a
function of line of sight is taken to be the fit shown in Figure \ref{fig:fsublos}, and biased cross 
section weighting assumes substructures do not significantly impact the lensing cross section.
For comparison, the mean substructure mass fraction for randomly oriented halos in our model
is $\avg{f_s}=0.46\%$.}
\label{fig:fsub}
\end{figure}

Using the fit to $f_s(\cos(\theta_{long}))$ shown in Figure \ref{fig:fsublos}, we compute the mean
projected substructure mass fraction $\avg{f_s|q_1,q_2}$ for a population of double and quad lenses 
as a function of the halo axis ratios $q_1$ and $q_2$.  Our results are shown in Figure \ref{fig:fsub}.
For reference, the mean substructure mass fraction for randomly oriented halos obtained from
the fit shown in figure \ref{fig:fsublos} is $\avg{f_s}=0.46\%$.   As per our expectations, we find that
prolate (cigar-like) doubles have substructure mass fractions that are  enhanced relative
to the average halo, with $\avg{f_s}\approx 0.6\%$.  Note though that this enhancement is relatively
minor, and slowly decreases to the random average as halos become oblate.


More interesting to us is the behavior of quads, for which we find a mild enhancement relative to random
for oblate halos, and a decrease in the expected substructure mass fraction for prolate halos.  This  
can be easily understood from Figure \ref{fig:losdist}: oblate quads strongly avoid projections along the short
axis of the halo, and projections along the middle and long axis of the lens are nearly equally likely.
Consequently, one expects an enhancement of the substructure mass fraction because some lines of sight with
low $f_s$ are avoided.  On the other hand, for prolate halos, projections along the long axis of the lens are
the least common, so indeed we expect the mean projected substructure mass fraction for these systems 
to be reduced.

Overall, though, it is clear that for quad systems - which are the only kind of systems for which $\avg{f_s}$ may be
estimated using the methods of \citet[][]{dalalkochanek02a} - the substructure mass fraction in the inner regions
of a halo cannot be significantly enhanced due to lens biasing if the impact of substructures on the lensing
cross section of galactic halos can be neglected.  Thus, lens biasing does little to soften the slight
(and in these authors' opinion, not terribly significant) discrepancy between the values of $f_s$ recovered 
by \citet[][]{dalalkochanek02a} and those from numerical simulations.


\section{Caveats and Systematics}
\label{sec:caveats}

Before we finish, we believe it is important to mention two systematics that could significantly
affect the conclusions presented in this work.  Specifically, throughout we have assumed
that the lensing cross section is dominated by the smooth mass distribution of lensing galaxies,
and we presented in section \ref{sec:model} several studies that suggest that our model
for the mass distribution of early type galaxies is a reasonable one.  As mentioned in
the introduction, the possible discrepancy between theory and observation concerning
the quad-to-double ratio of the CLASS lenses has raised the possibility that lensing
cross sections are in fact heavily influenced by the environment of the halo or
possibly by substructures with in it.  We briefly discuss each of these in turn.

We begin by discussing halo environments.  In our calculations above, and in
most of the lensing statistics literature, the effect of halo environment on lensing statistics is 
neglected.  This is not an entirely ad hoc assumption.  Theoretical estimates of 
the amount of shear that the typical lens experiences are quite small
\citep[$\gamma \approx 0.02$, see e.g.][]{keetonetal97, dalalwatson04}, so its impact should be negligible.  
Curiously, however, explicit lens modeling of known systems usually requires large external shears
($\gamma\approx 0.1$) in order to provide reasonable fits to observations \citep[see e.g.][]{keetonetal97}.
Moreover, direct estimates of the environment of lensing galaxies also support the idea of a stronger
effect from nearby structures \citep[][]{ogurietal05}.   The discrepancy between these
observations and the predictions for halo environments are themselves an interesting problem,
which ultimately may or may not be related to the usual quad-to-double ratio problem.  At any
rate, one might hope that even if such large external shears are correct, their impact on
the quad-to-double ratio would still be negligible if
their orientation is random.  This expectation was indeed confirmed by \citep[][]{rusintegmark01}.
Unfortunately, it is known that a significant fraction of lenses are actually member galaxies of intermediate
mass groups \citep[][]{momchevaetal06,williamsetal06}, and that galaxies in groups and clusters tend to be radially
aligned \citep[][]{pereirakuhn05, donosoetal06, faltenbacheretal07}, implying 
the randomly oriented shear assumption is likely not justified.
Indeed, a careful analysis of the impact of the halo environment for group members shows neglecting
to take said environment into account can lead to an underestimate of the ratio of the quad-to-double
lensing cross sections for such galaxies as large as a factor of two \citep[][]{keetonzabludoff04}.  
At this point, what seems clear is that there is not as of yet a definitive answer as to exactly how
important galaxy environments are, and thus, we have opted for making the simplest possible assumption
for the purposes of this work, that is, we have ignored the impact of large-scale environments.

The second solution to the quad-to-double ratio problem involves substructures.  Specifically, 
\citet[][]{cohnkochanek04}
have shown that the lensing cross section of galaxies is severely affected
by substructures.  If this is indeed the case, the way in which lensing galaxies are biased relative to
the overall galaxy population depend not only on the smooth component of its mass distribution, but
also on the spatial distribution of substructures within the galaxy halo.   Interestingly, in such a scenario
halo triaxiality would impact the orientation of halos relative to the line of sight now only through the biasing
due to the smooth matter component, but also because of the previously mentioned alignment between
the substructure distributions and the smooth mass distributions.  We leave the question of exactly how 
such a population of halos would be biased to future work (Chen et al., in preparation).


\section{Summary and Conclusions}
\label{sec:summary}

The triaxial distribution of mass in galactic halos implies that the probability that a galaxy becomes a lens
is dependent on the relative orientation of the galaxy's major axis to the line of sight.   Consequently, a
subsample of randomly oriented galaxies that act as strong lenses will {\it not} be randomly oriented
in space.  The relative orientation and the strength of the alignment depends on the shape of the 
matter distribution, and on the type of lens under consideration: prolate doubles have a high probability
of being project along their long axis, whereas the distribution of oblate doubles is nearly isotropic.
Prolate quads are most often projected along their middle axis, though the degree to which alignment
occurs is not as strong as for prolate doubles.  Interestingly, highly prolate quads are also more likely
to be projected along their short axis than along the long axis, though this very quickly changes as halos
become more triaxial and less prolate.  Oblate quads strongly avoid projections along the short axis of the lens,
but projections along the other two axis are almost equally likely.

An important consequence of the differences in the distribution of halo orientations for 
quad lenses, double lenses, and the galaxy population as a whole is that
the ellipticity distribution of these various samples must be different, even if the distribution of halo
shapes is the same.   Specifically, we predict that quad lenses are typically 
more elliptical than random galaxies, and that the ellipticity distribution of doubles is very slightly more circular
than that of random galaxies.  While current data do not show any indication of these trends,
we have shown that $\approx 300\ (1,400)$ lenses are necessary to obtain a $2\sigma\ (5\sigma)$ 
detection of the effect.  

The fact that halo triaxiality affects the ellipticity distribution of lensing galaxies also means that
halo triaxiality needs to be properly taken into account in lensing statistics.   Consequently,
we estimate how the biased lensing cross sections of galaxies depend on halo shape, and
find that they are nearly independent of the halo shape parameter $T$.  Instead, the mean
biased cross section of a lens depends almost exclusive on the distribution on the short-to-long
axis ratio $q_2$ (often denoted by $s$).  

Finally, given that the
distribution of substructures in numerical simulations is observed to be preferentially aligned with
the long axis of the host halos, we estimate how the preferred orientation of lensing galaxies
affects their predicted substructure mass fraction.  We find that biases due to non-isotropic distribution
of halos relative to the line of sight have an insignificant impact on the mean substructure mass fraction
of lensing galaxies.  

{\bf Acknowledgements: } ER would like to thank Christopher Kochanek for numerous discussions
and valuable comments on the manuscript 
which have greatly improved both the form and content of this work.  The authors
would also like to thank to Emilio Falco for kindly providing the isophotal axis ratio data that was needed 
for producing Figure \ref{fig:cumq}, and to Charles Keeton for a careful reading of the manuscript.
ER was funded by the Center for Cosmology and Astro-Particle Physics
(CCAPP) at The Ohio State University.  ARZ has been funded by the University of Pittsburgh, the National Science
Foundation (NSF) Astronomy and Astrophysics Postdoctoral Fellowship program through grant AST 0602122, and
by the Kavli Institute for Cosmological Physics at The University of Chicago.  This work made use of the National 
Aeronautics and Space Administration Astrophysics Data System.

\bibliographystyle{apj}
\bibliography{mybib}

\newcommand\AAA[3]{{A\& A} {\bf #1}, #2 (#3)}
\newcommand\PhysRep[3]{{Physics Reports} {\bf #1}, #2 (#3)}
\newcommand\ApJ[3]{ {ApJ} {\bf #1}, #2 (#3) }
\newcommand\PhysRevD[3]{ {Phys. Rev. D} {\bf #1}, #2 (#3) }
\newcommand\PhysRevLet[3]{ {Physics Review Letters} {\bf #1}, #2 (#3) }
\newcommand\MNRAS[3]{{MNRAS} {\bf #1}, #2 (#3)}
\newcommand\PhysLet[3]{{Physics Letters} {\bf B#1}, #2 (#3)}
\newcommand\AJ[3]{ {AJ} {\bf #1}, #2 (#3) }
\newcommand\aph{astro-ph/}
\newcommand\AREVAA[3]{{Ann. Rev. A.\& A.} {\bf #1}, #2 (#3)}

\appendix

\section{Lensing Cross Sections of Singular Isothermal Ellipsoids}

The Singular Isothermal Ellipsoid (SIE) is one of the simplest lens models that
can produce quadruply imaged sources.  \citet[][]{kormannetal94} performed a detailed 
study of the lensing properties
of SIE lenses, and, in particular, derived simple expressions for the total
area contained within the tangential and radial caustics of such lenses.
Specifically, given an SIE profile
\begin{equation}
\Sigma = \frac{\sqrt{q}\sigma_v^2}{2G}\frac{1}{x^2+q^2y^2},
\end{equation}
\citet[][]{kormannetal94} found that the area $\sigma_r$ and $\sigma_t$ contained 
inside the radial and tangential caustics is given by 
\begin{equation}
\sigma_r = \frac{4q}{1-q^2}\int_q^1dx\ \frac{\cos^{-1}(x)}{\sqrt{x^2-q^2}}
\end{equation}
and
\begin{equation}
\sigma_t = \frac{4q}{1-q^2}\int_q^1dx\ \left(\frac{\sqrt{1-x^2}}{x}-\cos^{-1}(x)\right)
	\frac{\sqrt{x^2-q^2}}{x^2}
\end{equation}
respectively.  Moreover, they showed that for $q>q_c$ where $q_c\approx 0.394$,
the tangential caustic is entirely contained within the radial caustic, and hence
the lensing cross section $\squad$ for forming four image lenses is simply
$\squad = \sigma_t$.  Likewise, the lensing cross section for forming doubles
is given by $\sdouble = \sigma_r-\sigma_t$.

Unfortunately, as derived in section \ref{sec:cs}, the relevant quantity for
lensing statistics of a flux limited sample is not the lensing cross section
itself, but the biased cross section $\sigma_B$.  Moreover, the latter cross
section requires one to compute
the magnification distribution $p(\mu)$ for double and quad lenses, for which
there are no closed form expressions.  In this appendix, we numerically compute
the magnification distribution $p(\mu)$, and its first moment $\avg{\mu}$ for both
doubles and quads, and use them to compute the biased lensing cross section
$\sigma_B=\avg{\mu}\sigma$ appropriate for a source luminosity function 
$n_s(L)\propto L^{-2}$.  

\begin{figure}[t]
\epsscale{1.15}
\plottwo{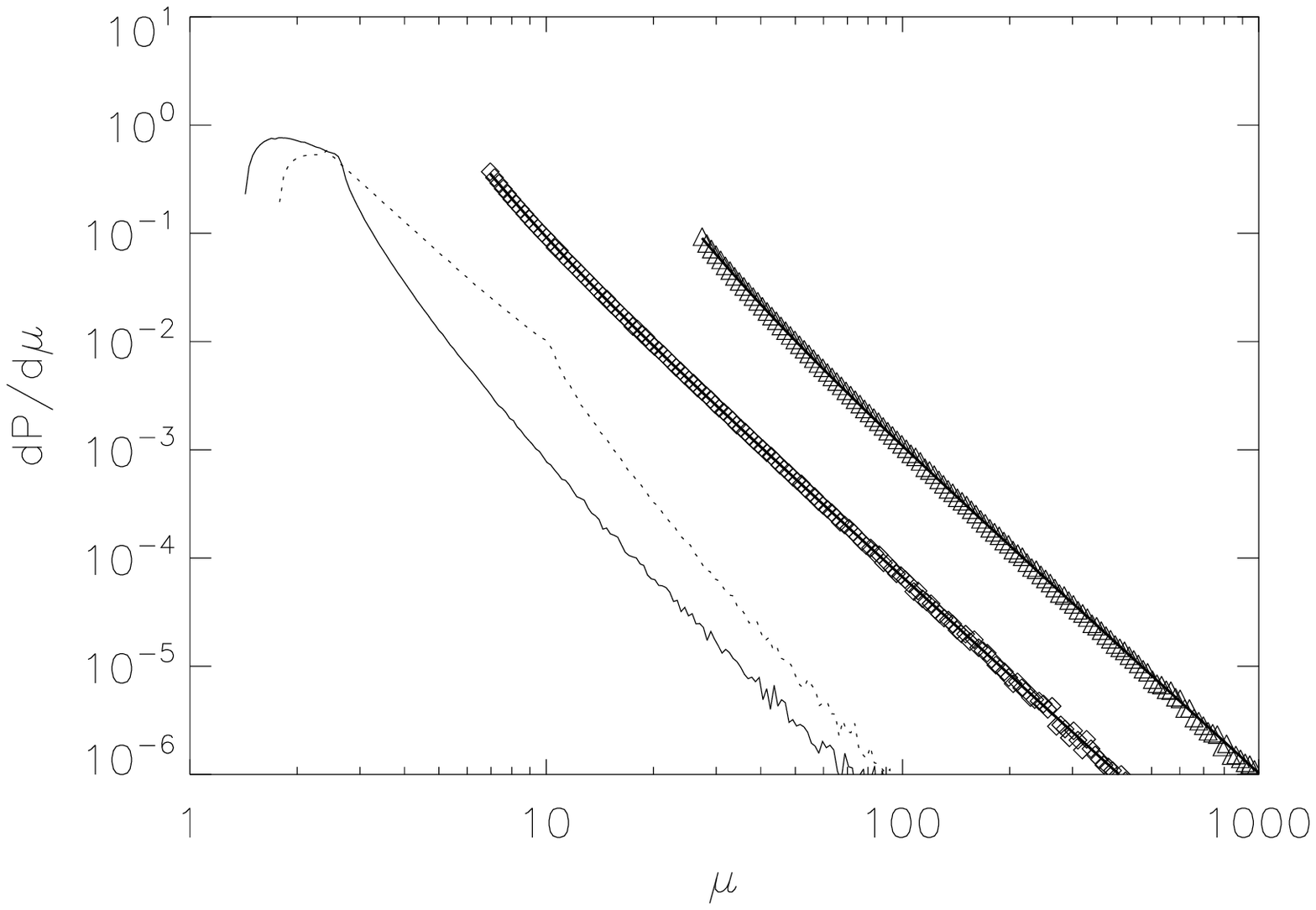}{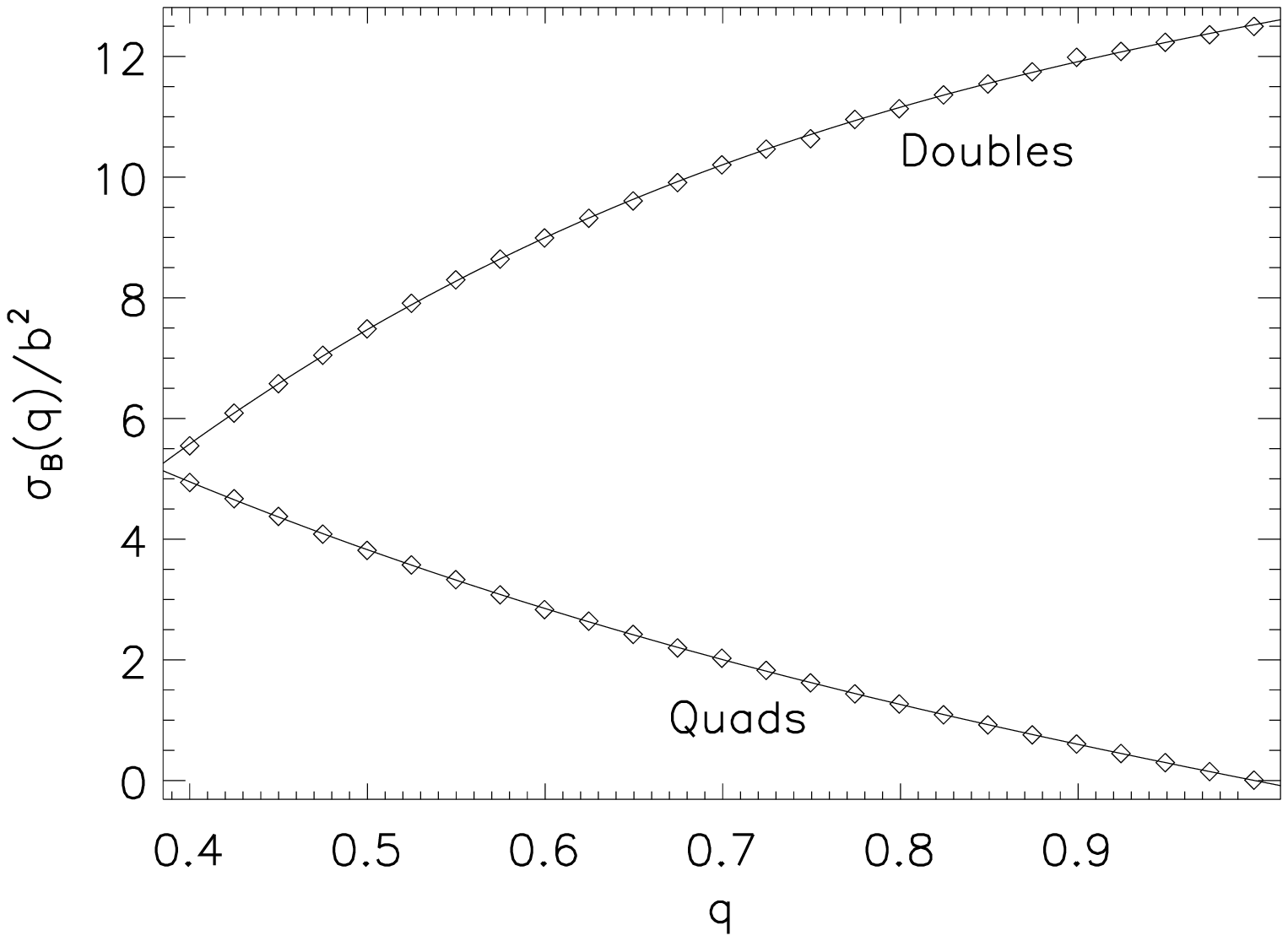}
\caption{{\it Left panel:} Distribution of magnifications $p(\mu)$ for double and quadruple lenses for
an SIE profile with axis ratios $q=0.4$ and $q=0.8$.  The $q=0.4$ distribution for doubles
is shown as a solid line, and the corresponding distribution for quads is shown as diamonds.
The dotted curve and triangles are the double and quad distributions for $q=0.8$.  The solid
curves on top of the diamonds and triangles are our empirical fit for quad lenses (see Eq.
\ref{eq:quadfit}, and is 
accurate to $\approx 5\%$.  {\it Right panel:} Biased lensing cross section $\sigma_B=\avg{\mu}\sigma$
as a function of axis ratio for SIE profiles and assuming a source luminosity function with a power law slope
$\alpha=-2$.  The solid curves are simple cubic fits which are accurate to better than $1\%$.}
\label{fig:pmu}
\end{figure}

The left panel of Figure \ref{fig:pmu} shows the magnification distribution for doubly and quadruply
image systems for SIE profiles with axis ratios $q=0.4$ and $q=0.8$.   Note that the magnification
distribution for doubles is very rich in features.  The magnification distribution for quads, on the other
hand, is relatively simple, and we can provide a simple fitting formula for it.  To do so, first note that we
know that in the limit $\mu\rightarrow \infty$, 
$\pquad \propto \mu^{-3}$, so we expect that $\pquad \approx Nx^{-3}f(x)$ where
$x=\mu/\mu_{min}$ and $\mu_{min}$ is the minimum magnification for quad 
lenses, $N$ is a normalization constant, and $f(x)$ is a function which asymptotes
to unity and deviates from unity only for $x\approx 1$.  Consequently, we expand
$f(x)$ in a power series in terms of $x^{-1}$, of which we expect only the first few terms
would be necessary to produce a good fit.  As it turns out, we found that $f(x)$
needs only one non-constant term to result in excellent fits to $\pquad$, and our
final fitting function for $\pquad$ is thus
\begin{equation}
\pquad \approx  \frac{1}{\mu_{min}^{(4)}} \frac{2}{1+a/2} x^{-3}(1+ax^{-2}).
\label{eq:quadfit}
\end{equation}
A priori, we would expect that the best fit value of the $a$ coefficient in the above
expression would be a function of the axis ratio $q$ of the profile.  While there does
appear to be some such dependence, it is extremely mild, so we have opted
for keeping $a$ fixed to the value $a=0.83$.    We found that this expression is 
accurate to better than $5\%$ for 
$\mu_{min} \lesssim \mu \lesssim 20\mu_{min}$ and $q\geq0.4$.


The right panel of Figure \ref{fig:pmu} shows the actual quantities we are interested in, the biased lensing
cross sections $\sigma_B = \avg{\mu}\sigma$.  As is obvious from the figure, the form of these biased 
cross sections is very simple, so even a simple quadratic fit results in quite good fits (of order a few percent).
Since we wish our empirical fit to be accurate, we fit the numerically computed cross sections with a cubic, 
which is enough to obtain sub-percent level accuracy.  Our best fit curves (in a least square sense) are
\begin{eqnarray}
\sigma_B^{(2)} & = & -6.902+42.937q-33.240q^2+9.736q^3 \\
\sigma_B^{(4)} & = &\ 11.409-20.833q+13.236q^2-3.816q^3.
\end{eqnarray}
Of course, we could have just as easily splined the numerically estimated values to compute the lensing
cross section at any axis ratio $q$.  We opted to fit the cross sections with a simple form both for simplicity,
and in the chance that the fitting formulae provided here will be useful for other works.

\section{Projected Surface Density Profiles of Triaxial Isothermal Halos}
\label{app:proj}

Consider an SIS profile
\begin{equation}
\rho_{SIS}(r) = \frac{\sigma_v^2}{2\pi G}\frac{1}{r^2}.
\end{equation}
Its triaxial generalization takes the form
\begin{equation}
\rho_{SIE}(\bar\bm{x}) = N(q_1,q_2)\frac{\sigma_v^2}{2\pi G}\frac{1}{\bar x^2/q_1^2+\bar y^2+\bar z^2/q_2^2}
\end{equation}
where $q_1$ and $q_2$ are the halo's axis ratios, and we have chosen a coordinate system
that is aligned the halo's principal axis and such that $1\geq q_1 \geq q_2$.  $p_2$ remains the ratio of the small to large axis. The prefactor $N(q_1,q_2)$ represents a relative normalization
for halos of varying axis ratios which we will compute shortly.  First however, 
since \citet[][]{kormannetal94} use the notation where the axis ratios multiply rather than divide the coordinates,
we rewrite the mass density as
\begin{equation}
\rho_{SIE}(\bar\bm{x}) = \tilde N(q_1,q_2)\frac{\sigma_v^2}{2\pi G}\frac{1}{p_1^2\bar x^2+p_2^2\bar y^2+\bar z^2}
\label{eq:triaxial}
\end{equation}
where $p_1=q_2/q_1$, $p_2=q_2$, and $\tilde N(p_1,p_2) = q_2^2N(q_1,q_2)$.  
Note $p_1$ is the ratio of the small to middle axis, while
$p_2$ remains the ratio of the small to large axis.
We choose the normalization function $\tilde N(p_1,p_2)$ such
that the mass contained within a radius $r$ is independent of the axis ratios $p_1$ and $p_2$,
as appropriate if one wishes to investigate the impact of triaxiality on lensing cross sections
at fixed mass with the latter defined using spherical overdensities. 
Integrating the above profiles and setting $M_{SIS}(r)=M_{SIE}(r)$ results in\footnote{To obtain the
expressions above, we perform first the radial integral and then the $\theta$ integral where $\theta$
is the azimuthal angle.}
\begin{equation}
\tilde N(p_1,p_2) = \left\{ \frac{2}{\pi} \int_{0}^{\pi/2} d\phi\ 
	\frac{\tan^{-1}\left[ \sqrt{(1-a)/a} \right]}{\sqrt{a(1-a)}}\right\}^{-1}
\end{equation}	
where we have defined $a(\phi;q_1,q_2)$ via
\begin{equation}
a(\phi;q_1,q_2) = p_1^2\cos^2(\phi)+p_2^2\sin^2(\phi).
\end{equation}

We wish to project $\rho_{SIE}$ along an arbitrary line of sight.  Let $\bm{x}$ be a coordinate system such
that the $z$ axis is aligned with the line of sight.  We choose the $x$ and $y$ axis to be such that a 
rotation by an angle $\theta$ along the $y$ axis followed by a rotation along the $z$ axis by an angle
$\phi$ recovers the coordinate system $\bar\bm{x}$ from Eq. \ref{eq:triaxial}.  The corresponding
rotation matrix is given by
\begin{equation}
R = \left(\begin{array}{ccc}
	\cos\theta\cos\phi & -\sin\phi & \sin\theta\cos\phi \\
	\cos\theta\sin\phi & \cos\phi & \sin\theta\sin\phi \\
	-\sin\theta & 0 & \cos\theta
	\end{array}\right).
\end{equation}

By construction, the corresponding projected surface density $\Sigma(x,y)$  is
given simply by
\begin{equation}
\Sigma(x,y) = \int_{-\infty}^\infty dz\ \rho_{SIE}(R\bm{x})
\end{equation}
which has the form
\begin{equation}
\Sigma(x,y) = \tilde N(p_1,p_2) \frac{\sigma_v^2}{2\pi G} \int_{-\infty}^\infty dz\ \frac{1}{A+Bz+Cz^2}
\label{eq:projint}
\end{equation}
where
\begin{eqnarray}
A & = & A_{xx}x^2+A_{xy}xy+A_{yy}y^2 \\
B & = & B_xx+B_yy \\
C & = & p_1^2\sin^2\theta\cos^2\phi+p_2^2\sin^2\theta\sin^2\phi+\cos^2\theta
\end{eqnarray}
and
\begin{eqnarray}
A_{xx} & = & p_1^2\cos^2\theta\cos^2\phi+p_2^2\cos^2\theta\sin^2\phi+\sin^2\theta \\
A_{xy} & = & \sin(2\phi)\cos(\theta)(-p_1^2+p_2^2) \\ 
A_{yy} & = & p_1^2\sin^2\phi+p_2^2\cos^2\phi \\ 
B_x & = & \sin(2\theta)(p_1^2\cos^2\phi+p_2^2\sin^2\phi-1) \\
B_y & = & \sin(\theta)\sin(2\phi)(-p_1^2+p_2^2).
\end{eqnarray}
Note that if $q_1=q_2=1$, then $A_{xx}=A_{yy}=C=1$ and $A_{xy}=B_x=B_y=0$, exactly as it should.
Performing the integral in Eq. \ref{eq:projint} we find
\begin{equation}
\Sigma(x,y) = \tilde N(p_1,p_2) \frac{\sigma_v^2}{2G}\frac{1}{\sqrt{AC-B^2/4}}
\end{equation} 
which has the generic form
\begin{equation}
\Sigma(x,y) = \tilde N(p_1,p_2) \frac{\sigma_v^2}{2G}\frac{1}{(\alpha_{xx}x^2+\alpha_{xy}xy+\alpha_{yy}y^2)^{1/2}}
\label{eq:projden}
\end{equation}
where
\begin{eqnarray}
\alpha_{xx} & = & A_{xx}C-B_x^2/4 \\
\alpha_{xy} & = & A_{xy}C-B_xB_y/2 \\
\alpha_{yy} & = & A_{yy}C-B_y^2/4.
\end{eqnarray}
For $q_1=q_2=1$, the above expressions reduce to $\alpha_{xx}=\alpha_{yy}=1$ and $\alpha_{xy}=0$
as appropriate for an SIS profile.  For the more general case
it is evident from equationuation \ref{eq:projden} that using an additional rotation of the $x-y$ plane 
we can diagonalize the projected mass density $\Sigma(x,y)$.  We find that the required 
rotation angle $\psi$ is given by
\begin{equation}
\tan2\psi = \frac{\alpha_{xy}}{\alpha_{xx}-\alpha_{yy}}.
\end{equation}
Using a $\sim$ to denote the new coordinate system, we can thus write
\begin{equation}
\Sigma(\tilde x, \tilde y) = \frac{\sqrt{q}\tilde \sigma_v^2}{2G}\frac{1}{(\tilde x^2+q^2\tilde y^2)^{1/2}}
\label{eq:projden1}
\end{equation}
where
\begin{eqnarray}
q^2 & = & \frac{\tilde\alpha_{yy}}{\tilde\alpha_{xx}} \label{eq:qproj} \\
\tilde\sigma_v^2 & = & \frac{\tilde N(p_1,p_2)}{\sqrt{q\tilde\alpha_{xx}}}\sigma_v^2 \label{eq:norm}
\end{eqnarray}
and we have defined
\begin{eqnarray}
\tilde\alpha_{xx} & = &\alpha_{xx}\cos^2\psi+\alpha_{xy}\sin\psi\cos\psi+\alpha_{yy}\sin^2\psi \\
\tilde\alpha_{yy} & = & \alpha_{xx}\sin^2\psi-\alpha_{xy}\sin\psi\cos\psi+\alpha_{yy}\cos^2\psi.
\end{eqnarray}
As expected, the above expression for $q$ reduces to $q=p_2/p_1=q_1$ when we project along the
$z$ axis (i.e. the short axis), to $q=p_2$ when projecting along the $y$ axis (i.e. the long axis), 
and to $q=p_1=q_2/q_1$ when projecting along the $x$ axis (i.e. the middle axis).  
The particular form of the parameterization of the surface density in Eq. \ref{eq:projden1}
is meant to match the conventions in \citet[][]{kormannetal94}, which was chosen to ensure 
the mass contained within a given density contour be independent of $q$ for fixed $\tilde\sigma_v$.

and the

\end{document}